\documentclass[12pt,dvips]{article}
\usepackage[OT1]{fontenc}
\usepackage{graphicx,color,pstricks}
\usepackage{fancybox,amsmath,amssymb,epsfig,rotating}
\usepackage{color,array,shadow}
\def\chain#1#2#3{{\relax\hbox{
        \raise 40pt\vtop{
               \begin{tabbing}
                       $#1{}$\=$#2$\\
                       \>$\; |$\\ [-9pt]
                       \>$\ \rightarrow #3$\\
               \end{tabbing}           }}}}

\begin{document}
\newcommand{\ra}        {\mbox{$\rightarrow$}}
\newcommand{\bc}        {\begin{center}}
\newcommand{\ec}        {\end{center}}
\newcommand{\be}        {\begin{equation}}
\newcommand{\ee}        {\end{equation}}
\newcommand{\gam}       {\mbox{$\gamma$}}
\newcommand{\piz}       {\mbox{$\pi^0$}}
\newcommand{\pip}       {\mbox{$\pi^+$}}
\newcommand{\pim}       {\mbox{$\pi^-$}}
\newcommand{\etg}       {\mbox{$\eta$}}
\newcommand{\etp}       {\mbox{$\eta^{\prime}$}}
\newcommand{\omg}       {\mbox{$\omega$}}
\newcommand{\rh}        {\mbox{$\rho$}}
\newcommand{\pbp}       {\mbox{$\bar{p}p$}}
\newcommand{\ssb}       {\mbox{$\bar{s}s$}}
\newcommand{\NNb}       {\mbox{$\overline{N}N$}}
\newcommand{\pbar}       {\mbox{$\bar{p}$}}

\definecolor{lightyellow}{cmyk}{0,0,0.5,0}
\definecolor{lightred}{rgb}{1,0.5,0.5}
\definecolor{lightgreen}{rgb}{0.5,1,0.5}
\definecolor{lightblue}{rgb}{0.5,0.5,1}
\definecolor{darkgreen}{rgb}{0,0.5,0}
\definecolor{darkcyan}{cmyk}{1,0.3,0.3,0.3}
\definecolor{darkblue}{rgb}{0.5,0.5,1}
\definecolor{lightbrown}{rgb}{0.7,0.3,0.3}
\definecolor{darkbrown}{rgb}{0.5,0,0}
\definecolor{bluegreen}{rgb}{0,0.5,0.5}
\def\chain#1#2#3{{\relax\hbox{
        \raise 40pt\vtop{
               \begin{tabbing}
                       $#1{}$\=$#2$\\
                       \>$\; |$\\ [-9pt]
                       \>$\ \rightarrow #3$\\
               \end{tabbing}           }}}}
\definecolor{fill1}{rgb}{0.5,0.0,0.0}
\definecolor{shadow1}{rgb}{0.25,0.0,0.0}
\definecolor{fill2}{rgb}{0.0,0.5,0.0}
\definecolor{shadow2}{rgb}{0.0,0.25,0.0}
\definecolor{fill3}{rgb}{0.34,0.0,0.72}
\definecolor{shadow3}{rgb}{0.17,0.0,0.36}
\definecolor{fill_gl}{rgb}{0.9,0.9,0.9}
\definecolor{shadow_gl}{rgb}{0.45,0.45,0.45}
\newcommand{\nonett}[9]
{
\setlength{\unitlength}{1mm}
\begin{picture}(150.00,90.00)
\put(10.00,45.00){\vector(1,0){70.00}}
\put(45.00,10.00){\vector(0,1){70.00}}
\put(110.00,45.00){\vector(1,0){30.00}}
\put(125.00,30.00){\vector(0,1){30.00}}
\put(82.50,42.50){\makebox(5.00,5.00){$I_3$}}
\put(142.50,42.50){\makebox(5.00,5.00){$I_3$}}
\put(127.50,60.00){\makebox(5.00,5.00){S\quad\ Singlet }}
\put(47.50,80.00){\makebox(5.00,5.00){S\quad\ Octet }}
\put(45.00,45.00){\circle*{2.00}}
\put(70.00,45.00){\circle*{2.00}}
\put(20.00,45.00){\circle*{2.00}}
\put(32.50,70.00){\circle*{2.00}}
\put(57.50,70.00){\circle*{2.00}}
\put(32.50,20.00){\circle*{2.00}}
\put(57.50,20.00){\circle*{2.00}}
\put(125.00,45.00){\circle*{2.00}}
\put(45.00,45.00){\circle{0.00}}
\put(45.00,45.00){\circle{5.00}}
\put(32.50,70.00){\line(1,0){25.00}}
\put(57.50,70.00){\line(1,-2){12.50}}
\put(70.00,45.00){\line(-1,-2){12.50}}
\put(57.50,20.00){\line(-1,0){25.00}}
\put(32.50,20.00){\line(-1,2){12.50}}
\put(20.00,45.00){\line(1,2){12.50}}
\put(60.00,70.00){\makebox(15.00,5.00)[l]{#2}}
\put(15.00,70.00){\makebox(15.00,5.00)[r]{#1}}
\put(15.00,15.00){\makebox(15.00,5.00)[r]{#3}}
\put(6.75,37.50){\makebox(13.25,5.00)[r]{#5}}
\put(60.00,15.00){\makebox(15.00,5.00)[l]{#4}}
\put(70.00,37.50){\makebox(13.75,5.00)[l]{#6}}
\put(47.50,37.50){\makebox(12.50,5.00)[l]{#7}}
\put(47.50,47.50){\makebox(12.50,5.00)[l]{#8}}
\put(127.50,47.50){\makebox(12.50,5.00)[l]{#9}}
\end{picture}
}

%%%%%%%%%%%%%%%%%%%%%%%%%
%
% End of definitions
%
%%%%%%%%%%%%%%%%%%%%%%%%%
%
%
% text defaults
%
\newcommand{\Ra}{$\bf\Rightarrow$}
\renewcommand{\bottomfraction}{0.9}
\renewcommand{\topfraction}{0.9}
\renewcommand{\textfraction}{0.1}
%%%%%%%%%%%%%%%%%%%%%%%%%%%%%%%%%%%%%%%%%%%%%%%%%%%%%%%%%%%%%%%%%%%%%%%%%%%%
%%
%%%%%%%%%%%%%%%%%%%%%%%%%%%%%%%%%%%%%%%%%%%%%%%%%%%%%%%%%%%%%%%%%%%%%%%%%%%%
%%
%%%%%%%%%%%%%%%%%%%%%%%%%%%%%%%%%%%%%%%%%%%%%%%%%%%%%%%%%%%%
%               title page
%%%%%%%%%%%%%%%%%%%%%%%%%%%%%%%%%%%%%%%%%%%%%%%%%%%%%%%%%%%%
%
\title{Baryon resonances and strong QCD }

\vskip 30mm
\author{E. Klempt
\\
Institut f\"ur Strahlen -- und Kernphysik, \\
Universit\"at Bonn,\\
D--53115 Bonn }
\maketitle
{\it Abstract:
Light-baryon resonances (with u,d, and s quarks in the SU(3)
classification) fall on Regge trajectories. When their squared masses
are plotted against the intrinsic orbital angular momenta {\rm L}, 
$\Delta^*$'s with even and odd parity can be described by the same
Regge trajectory. For a given {\rm L}, nucleon resonances with spin 
{\rm S}=3/2 are approximately degenerate in mass with $\Delta$ 
resonances. To which total angular momentum {\rm L} and {\rm S} 
couple has no significant impact
on the baryon mass. Nucleons with spin 1/2 are shifted in mass; the
shift is - in units of squared masses - proportional to the component
in the wave function which is antisymmetric in spin and flavor. Based
on these observations, a new baryon mass formula is proposed which
reproduces nearly all known baryon masses. It is shown that the masses
are compatible with a quark-diquark picture while the richness of the
experimentally known 
states require three particles to participate in the dynamics. 
This conflict is resolved by proposing that quarks polarize
the QCD condensates and are surrounded by a polarization cloud 
shielding the color. A new interpretation of constituent quarks as
colored quark clusters emerges; their interaction is responsible for
the mass spectrum. Fast flavor exchange between the colored quark
clusters exhausts the dynamical richness of the three-particle
dynamics. 
\par
The colored-quark-cluster model provides a mechanism in which the
linear confinement potential can be traced to the increase of the
volume in which the condensates are polarized. The quark-spin magnetic
moment induces currents in the polarized condensates which absorb
the quark-spin angular momentum: the proton spin is not carried by 
quark spins. The model provides a new picture of hybrids and
glueballs. 
}
\vskip 10mm

\clearpage
%\tableofcontents
%\clearpage
\section{Introduction}
Baryon spectroscopy has played a decisive role in the development of the quark
model and of flavor SU(3). The prediction of the $\Omega^-$
carrying total strangeness $S=-3$ \cite{Gell-Mann:nj} and its 
subsequent experimental discovery at the anticipated mass \cite{Barnes:pd}
was a triumph of SU(3). From the demand that the baryon wave function
be antisymmetric with respect to the exchange of two quarks, the need
of a further quark property was deduced \cite{Greenberg:pe}
which later was called color \cite{Gell-Mann:ph} and 
found to play an eminent dynamical role. The linear
dependence of the 
squared masses of baryons on their total angular momentum led to
the Regge theory of complex angular momenta \cite{Regge:mz}. 
The unsuccessful attempts
to 'ionize' protons and to observe free quarks \cite{Lyons:1984pw}
was the basis for the confinement hypothesis \cite{Wilson:1974sk}.
\par
In this paper a phenomenological analysis of the spectrum of light
baryons is presented. The three quarks with flavor up, down and 
strange may combine to the ground state octet baryons with total 
angular momentum J=1/2 or to the decuplet carrying angular momentum 3/2. 
These are the 'ground states' even though some of them  have large
hadronic widths: the $\Delta (1232)$, e.g., has a width of 120
MeV. Today, the Particle Data Group lists about 100 baryon
resonances, 85 of them have an experimentally determined spin and
parity, 50 baryon resonances are well established, having 3* or 4*
in the PDG notation \cite{Groom:in}. There 
is hence the hope, that a systematic investigation of the baryon
masses reveals the internal interactions between quarks in the
confinement region \cite{Bjorken:2000ni}.  
\par
The study presented here shows that the orbital angular
momentum is decisive for the excitation energies. 
The squared baryon masses depend linearly on the intrinsic orbital
angular momentum; spin-orbit and spin-spin splittings due to
color-magnetic forces are negligible. The well-known N-$\Delta$
splitting is traced to instanton-induced interactions. 
The observations can be condensed into a new baryon mass
formula containing four parameters. The mass formula reproduces most
known baryon masses with good accuracy. 
\par
The slope of the Regge trajectories of baryons is the same as the
mesonic Regge slope. In the 70'ties, baryons could be divided
into L-even baryons in SU(6) 56-plets, and L-odd baryons in 70-plets.
These observations supported the assumption that quark-diquark
interactions are responsible for the baryon masses
\cite{Anselmino:1992vg}. A SU(6) classification of the now-existing baryon
resonances will show that the mass spectrum is much richer and
reflects the full freedom of three-particle dynamics. 
\par
Now there is a clear-cut contradiction: the richness of the mass
spectrum evidences that three particles take place in the
interaction. This is of course what we would have anticipated. On the
other hand, the masses are all well described in a quark-diquark
picture. These experimental findings suggest a new
definition of 'constituent' quarks as colored clusters in which 
current-quarks polarize the $\rm q\bar q$ and gluon condensates of the QCD
vacuum. The color of the current quark is shielded by the
polarization cloud; the polarization cloud absorbs gluons before they
propagate to another current quark. The condensates themselves become
colored, as suggested by Wetterich and collaborators
\cite{Wetterich:2000pp}. The colored quark-clusters define
the interaction. Each quark experiences the forces exerted by a
colored diquark. The forces are hence equivalent to quark-antiquark
forces. Quark flavor is exchanged with a high frequency; to
first order, all SU(6) states of given intrinsic spin and orbital
angular momenta and with the same quark content are
degenerate in mass. Mass splitting occurs due to the spin- and 
flavor-dependence of instanton-induced interactions. 
\par
The new definition of a constituent quark has significant impact on
the comprehension of strong interactions. It suggests an interpretation of 
the confinement mechanism similar to that proposed by Nambu
\cite{Nambu:1978bd}.    
It provides a natural interpretation of the proton {\em spin crisis}: 
as often discussed, only a small fraction of the proton spin can be
ascribed to quarks (including the sea quarks) 
\cite{Filippone:2001ux}, the
largest contribution to the proton spin must be assigned to orbital
angular momenta of quarks and gluons, or to the gluon spin. In the new
picture proposed here, the quark spin induces magnetic currents in the
polarized condensates; the contribution of constituent quarks to the
nucleon spin is small. For gluonic excitations, hybrids and
glueballs, a new interpretation is proposed.  
\par
The outline of this paper is as follows: after a short introduction
into the ideas behind present quark-models of baryons, we give an
outline of SU(6) and of the symmetries of baryonic wave functions. In
section 4, phenomenological aspects of the light-baryon spectra are
presented. Regge trajectories are exploited to demonstrate the most
important forces responsible for the baryon mass pattern. Based on 
these results a new baryon mass formula is proposed 
(section \ref{sec_massformula}) and
compared to data. The comparison of experimental masses and
predictions of the model requires to assign baryons to SU(6)
multiplets; this assignment is discussed in section~6. 
\par 
The new mass formula requires a different concept of the constituent
quark and a new interpretation of the QCD forces. This aspect is
discussed in section 7. The paper ends with a short summary. 
\section{Baryon models}
It has been stressed very often
that the strong-interaction coupling-constant $\alpha_s$ increases
dramatically and diverges when the momentum transfer q approaches the QCD
scale parameter $\Lambda_{\rm QCD}$. This is the regime of strong QCD where
perturbative methods fail. Bound states of quarks, mesons and
baryons, involve very small momentum transfers; models need to be
developed to appreciate the meaning of strong interactions in this 
low-energy domain. 
\par
There are a few distinct classes of baryon models based on different
quark-quark and quark-antiquark interactions. Most models start from the
assumption that QCD generates a confinement
potential which grows linearly with the distances between the
quarks. The color-degrees-of-freedom guarantee the antisymmetry of the
baryon wave function. The equation of motion is solved after the
color-degrees-of-freedom have been integrated out: color plays no
dynamical role in the interaction. The confinement
potential corresponds to the mean potential energy experienced by a
quark at a given position, with a fast color exchange between the
three quarks. Confinement does not exhaust the full QCD
interaction: there are residual interactions which can be
parameterized in different ways. 
\par
The celebrated  
Isgur-and-Karl model starts from an effective spin-spin interaction
from one-gluon exchange the
strength of which is adjusted to match the $\Delta (1232)$--N mass 
difference. This requires a rather large value for 
$\alpha_s$, certainly invalidating a
perturbative approach. However, the one-gluon exchange is supposed to
sum over many gluonic exchanges which in total carry the quantum
numbers of a gluon. 
\par
Now there is an immediate problem: with this large
one-gluon exchange contribution, the spin-orbit splitting becomes very
large, in contrast to the experimental findings. Isgur and Karl solved
this problem by assuming that the Thomas precession in the confinement
potential leads to a spin-orbit splitting which cancels exactly the
spin-orbit coupling originating from one-gluon exchange.  This assumption
allowed to reproduce the low-lying baryon resonance masses and was a
break-through in the development of quark models for baryons \cite{IsgurKarl}.
Later, this model was further developed and
refined, relativistic corrections were applied and the full energy
spectrum of the relativized Hamiltonian was calculated. Results of the
latest variant of this type of model can be found in \cite{Capstick:bm}. 
\par
An alternative model was developed by Glossman and Riska
\cite{Riska}. The model is based on the assumption that pions or, more
generally,  Goldstone bosons are exchanged
between constituent quarks.  The phenomenological
success is impressing, in particular the low-lying P$_{11}$(1440), the
Roper resonance, is well reproduced. They emphasize the presence of
parity doublets which they believe to signal chiral-symmetry
restoration at high baryon masses.  
\par
The group of Metsch and Petry developed a relativistic
quark model with instanton induced two-body and three-body
interactions \cite{Metsch}.  The confinement forces - which in most models are
defined only in a non-relativistic frame and given as linear potential
in the three-particle rest frame - have a complex Lorentz structure. 
They solve the Bethe-Salpeter equation by reducing it to the
Salpeter equation. The parity doublets are naturally explained by
instanton interactions.
\par
Rijker, Iachello and Leviatan suggested an algebraic model
of baryon resonances \cite{Bijker:yr}. Their mass
formula is similar to the one proposed here, but uses 10 parameters
where most of them have no intuitive physical significance. On the
other hand, wave functions are constructed and transition amplitudes
can thus be calculated. 

\section{Symmetry considerations}
\label{class}
\par
\subsection{The baryonic wave function}
Symmetries play a decisive role in the classification of baryon
resonances. The baryon wave function 
can be decomposed into a color wave function, which is
antisymmetric with respect to the exchange of two quarks, the
spatial and the spin-flavor wave function. The second ket in the
wave function 
\begin{eqnarray}
\rm  |qqq> = |colour>_A \cdot &\rm |space;\ spin, flavour>_S
 \\
&\hspace*{-10mm}\rm O(6) \qquad SU(6) \nonumber
\end{eqnarray}
has to be symmetric. 
The SU(6) part can be decomposed into SU(3)$\otimes$SU(2).
\subsection{SU(3)}
In this paper, we restrict ourselves to light flavors, to $up, down$
and $strange$ quarks. The flavor wave function is then given by SU(3)
and allows a decomposition
\begin{equation}
\rm 3 \otimes\ 3 \otimes\ 3 = 10_S \oplus\ 8_M \oplus\ 8_M \oplus\ 1_A,
\end{equation}
into a decuplet which is symmetric w.r.t. the exchange of any two quarks,
a singlet which is antisymmetric and two octets of mixed symmetry. The
two octets have different SU(3) structures, only one of them
fulfills the symmetry requirements in the total wave functions.
Remember that the SU(3) multiplets contain six particle families:
\bc
\renewcommand{\arraystretch}{1.4}
\begin{tabular}{|ccccccc|}
\hline
SU(3) &  N  & $\Delta$ & $\Lambda$ & $\Sigma$ & $\Xi$ & $\Omega$ \\ 
 1    & no  &  no      &   yes     &   no     &   no  &    no    \\
 8    & yes &  no      &   yes     &   yes    &   yes &    no    \\
10    & no  &  yes     &   no      &   yes    &  yes  &    yes   \\
\hline
\end{tabular}
\renewcommand{\arraystretch}{1.0}
\ec

\subsection{SU(6)}
The spin-flavor wave function 
can be classified according to SU(6). 
\begin{equation}
\rm 6 \otimes\ 6 \otimes\ 6 = 56_S \oplus\ 70_M \oplus\ 70_M \oplus\ 20_A
\end{equation}
In the ground state, the spatial wave function is symmetric, 
and the spin-flavor wave function has to be symmetric, too. 
Then, spin and flavor can both be symmetric; this is the case
for the decuplet. Or spin and flavor wave function can individually
have mixed symmetry, with symmetry in the combined spin-flavor wave
function. This coupling represents the baryon octet. The 56-plet thus
decomposes into a decuplet with spin 3/2 (four spin projections) plus
an octet with
spin 1/2 (two spin projections) according to
\begin{equation}
56 = {^4}10\ \oplus\  ^{2}8.
\end{equation}
\par
The spin-flavor wave functions can also have mixed
symmetry. The 70-plet can be written as
\begin{equation}
70 = {^2}10\ \oplus\ {^4}8\ \oplus\ {^2}8\ \oplus\ {^2}1.
\end{equation}
Decuplet baryons, e.g. $\Delta^*$, in the 70-plet
have intrinsic spin
1/2; octet baryons like excited nucleons can have spin 1/2 or
3/2. Singlet baryons with J=1/2, the $\Lambda_1$ resonances, exist
only for spin-flavor wave functions of mixed symmetry. The ground
state (with no orbital excitation) has no $\Lambda_1$. 
\par
The 20-plet is completely antisymmetric and requires an antisymmetric
spatial wave functions. It is decomposed into an octet with spin 1/2
and a singlet with spin 3/2:
\begin{equation}
20 = {^2}8\ \oplus\ {^4}1.
\end{equation}
\subsection{The spatial wave function}
\par
The three-particle motion can be
decomposed, in Jacobian coordinates, into two relative motions and the
center-of-mass motion. The two relevant internal motions 
may support rotational and vibrational excitations leading to a large
number of expected resonances. The spatial wave functions of mesons
can be classified in the 3-dimensional rotational group O(3); the
three-body motion requires O(6). 
\par
The quark dynamics can be approximated by two harmonic oscillators; 
to first order, harmonic-oscillator wave-functions can be use. 
The rotational group O(6) can be expanded into 
O(6)$\rightarrow$O(3)$\otimes$O(2), see e.g. \cite{Hey:1982aj}. 
Table \ref{tab:hey} gives 
the expected multiplet structure in an O(6)$\otimes$SU(6)
classification scheme for the four lowest excitation 
quantum numbers $N$. With increasing $N$, an increasing number of
multiplets develop. The decomposition of the orbital wave-functions 
results in a complicated multiplet structure of harmonic-oscillator 
wave-functions. It should be mentioned that some of these multiplets need two
quark excitations. In the lowest 20-plet, at $N=2$, two quarks are
excited, each carrying one unit of orbital angular momentum; 
the two orbital angular momenta add to a total orbital angular
momentum 1 and positive parity. 
\par
The ground state $N=0$ is readily identified with the well-known octet
and decuplet baryons. The first excitation band ($N=1$) has internal
orbital angular momentum  L=1 ; both oscillators are excited
coherently, there is one coherent excitation mode of the two
oscillators. This information is comprised in the notation 
$3\otimes 2_1$. The next excitation band involves several dynamical
realizations. The intrinsic orbital angular momentum L can be
associated with 
two different quarks; the vector sum of the two l$_i$ can be 0, 1 or 2
giving rise to the series  $5\otimes 1$ to $1\otimes 1$ where the
$3\otimes 1$ is antisymmetric w.r.t. quark exchange, and the other
two are symmetric. 
Two linearly independent coherent two-oscillator excitations exist
which have mixed-symmetry spatial wave-functions. The baryon can
be excited radially where the three quarks oscillate against their
common center of mass. This mode is represented by $1\otimes 1$. 
\par
With increasing $N$, the number of multiplets increases strongly; 
multiplets belonging to bands of up to 12 were calculated
\cite{Dalitz:me}.
 The multitude of predicted resonances escaped so far
experimental observation. This is the  so-called  missing-resonance
problem, and the basis for experimental searches for new states
\cite{Napolitano:1993kf,Thoma-prop}. 
\par
\begin{table}[h!]
\caption{Multiplet-structure of harmonic oscillator wave functions. 
(From Hey and Kelly, Phys. Rep. 96 (1986) 71). The baryons are
classified into bands that have the same number {\em N} of excitation
quanta. 
D represents the dimension of the SU(6) representation, L and P
angular momentum and parity of the resonance, respectively.
\label{tab:hey}}
\bc
\renewcommand{\arraystretch}{1.3}
\begin{tabular}{|cccc|}
\hline
$N$&  O(6) &   O(3)$\otimes$O(2)   &   (D,L$^{\rm P}_{N}$)    \\
\hline
0  &   1   & $\ 1\otimes 1$        &  $\ (56,0^+_0)$  \\
1  &   6   & $\ 3\otimes 2_1$      &  $\ (70,1^-_1)$  \\
2  &  20   & $\ (5+1)\otimes 2_2$  &  $\ (70,2^+_2)$, $\ (70,0^+_2)$    \\
   &       & $\ 5\otimes 1$        &  $\ (56,2^+_2)$  \\      
   &       & $\ 3\otimes 1$        &  $\ (20,1^+_2)$  \\
   &   1   & $\ 1\otimes 1$        &  $\ (56,0^+_2)$  \\
3  &  50   & $\ (7+3)\otimes 2_3$  &  $\ (56,3^-_3)$, $\ (20,3^-_3)$, 
$\ (56,1^-_3)$, $\ (20,1^-_3)$ \\
   &       & $\ (7+5+3)\otimes 2_1$&  $\ (70,3^-_3)$, $\ (70,2^-_3)$, $\ (70,1^-_3)$ \\
   &   6   & $\ 3\otimes 2_1$      &  $\ (70,1^-_3)$  \\
4  & 105   & $\ (9+5+1)\otimes 2_4$&  $\ (70,4^+_4)$, $\ (70,2^+_4)$, $\ (70,0^+_4)$ \\
   &       & $\ (9+7+5+3)\otimes 2_2$&  $\ (70,4^+_4)$, $\ (70,3^+_4)$, 
$\ (70,2^+_4)$, $\ (70,1^+_4)$ \\
   &       & $\ (9+5+1)\otimes 1$  &  $\ (56,4^+_4)$, $\ (56,2^+_4)$, $\ (56,0^+_4)$ \\ 
   &       & $\ (7+5)\otimes 1$    &  $\ (20,3^+_4)$, $\ (20,2^+_4)$  \\  
   &  20   & $\ (5+1)\otimes 1$    &  $\ (70,2^+_4)$, $\ (70,0^+_4)$  \\
   &       & $\ 3\otimes 1$        &  $\ (20,1^+_4)$                \\
   &       & $\ 5\otimes 1$        &  $\ (56,2^+_4)$                \\
   &   1   & $\ 1\otimes 1$        &  $\ (56,0^+_4)$                \\
\hline
\end{tabular}
\renewcommand{\arraystretch}{1.0}
\ec
\end{table}

\clearpage
\section{Phenomenology}
\subsection{Regge trajectories, I}
\begin{figure}[h]
\bc
\epsfig{file=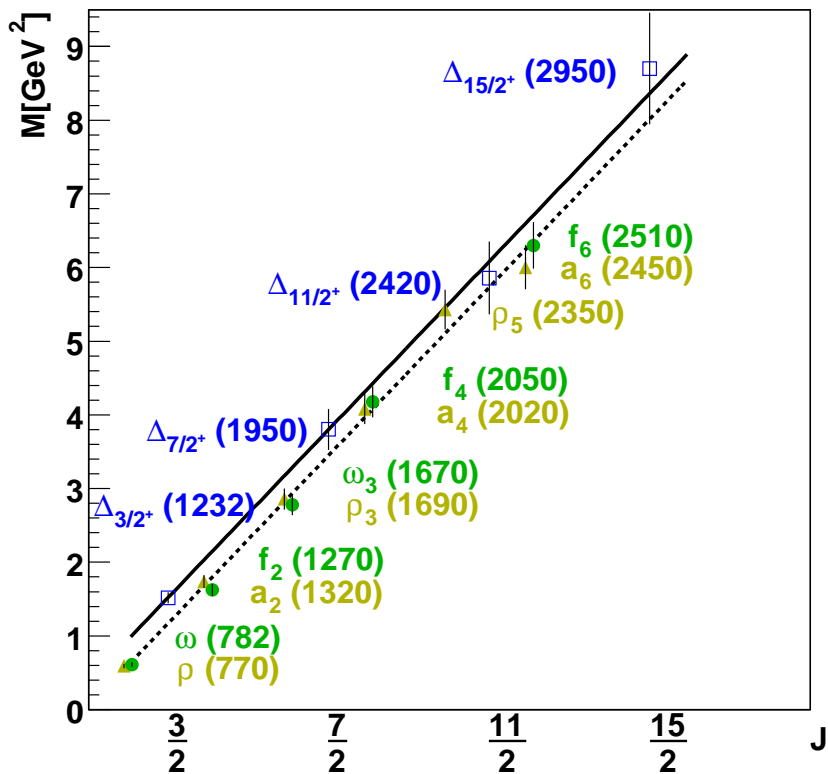,width=12cm}
\ec
\vspace*{-10mm}
\caption{$\Delta^*$'s with L even and J=L+3/2.  
Also shown is the Regge trajectory for mesons with
J=L+S. \label{Delta-mesons}}
\end{figure}

It is well known that meson and baryon resonances lie on Regge trajectories,
that their squared masses depend linearly on the total angular
momentum J. Fig. \ref{Delta-mesons} shows such a plot; $\Delta$
resonances are plotted having the lowest mass at a given total angular
momentum J, with J=L+3/2 and with orbital angular momentum L
even. The errors assigned will be discussed below.
\par
The Figure also shows a meson Regge trajectory, 
again as a function of the total angular momentum. Light mesons with 
approximate isospin degeneracy and with J=L+1 are presented. The
dotted line represents a fit to the 
meson masses taken from the PDG \cite{Groom:in}; the error in the fit is
given by the PDG errors and a second systematic error of 30 MeV added
quadratically. The slope is determined to  1.142 GeV$^2$. The 
$\Delta$ trajectory is given by the $\Delta (1232)$ mass and the slope
as determined from the meson trajectory. Obviously, mesons and $\Delta$'s 
have the same Regge slope. This observation is the basis for diquark models;
indeed, the QCD forces between quark and antiquark are the same as
those between quark and diquark. 
\par
\vspace*{-10mm}
\subsection{Spin-orbit coupling}
The starting point of our phenomenological discussion is the
observation that there are no or little spin-orbit splittings in the
baryon spectrum. The absence of spin-orbit splittings or, more
precisely, the smallness of its contribution to meson and baryon
resonances is a hotly debated subject \cite{Isgur-Riska}. 
Here, we discuss first the empirical facts and implications.
\par
\begin{figure}[h]
\vspace*{-10mm}
\bc
\epsfig{file=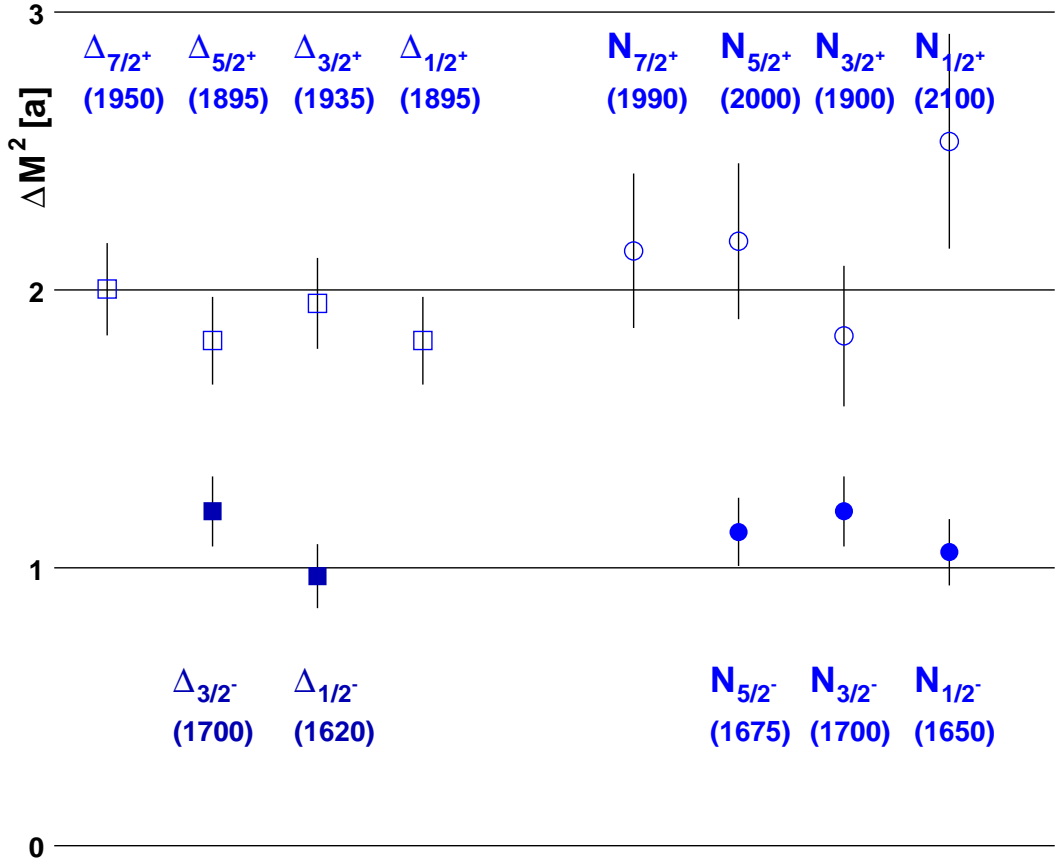,width=14cm}
\ec
\vspace*{-10mm}
\caption{$\Delta$ and N resonances assigned to super-multiplets with
defined spin and orbital angular momentum. Shown is the increase in mass 
square above 
the $\Delta (1232)$ [in units of a=1.142 GeV$^2$].
Upper panel: N$^*$ and $\Delta^*$
with L=2 and S=3/2 coupling to 
$\vec{\rm J} (7/2^+\,,5/2^+\,,3/2^+\,,1/2^+)$. Lower panel:
$\Delta^*$ with $\vec{\rm L} (1) + \vec{\rm S}  (1/2) = \vec{\rm J} 
(3/2^-\,,1/2^-)$ and N$^*$ with $\vec{\rm L} (1) + \vec{\rm S} (3/2) 
= \vec{\rm J} (5/2^-\,,3/2^-\,,1/2^-)$
In this and the following Figures, $\Delta$'s are represented by
squares, nucleons by circles. Open symbols characterize even, full
symbols odd parity. \label{ls-split}
}
\end{figure}
Fig. \ref{ls-split} shows squared masses of (selected) positive and
negative parity N and $\Delta$ resonances. The lines indicate the 
squared-mass values from the Regge trajectory at the first and second
excitation energy.  
In the upper panel, there are two groups of N and $\Delta$
resonances at 1.95 GeV, two super-multiplets, with 
$\rm J^P=7/2^+,5/2^+,3/2^+,1/2^+$. We assign intrinsic orbital angular
momentum L=2 and intrinsic spin S=3/2 to these states, with nearly
vanishing spin-orbit couplings. (As discussed below, the 
N$_{1/2^+}(2100)$ could also be the third radial excitation.)
Similarly we have, at 1650 MeV, 
two $\Delta$ states with L=1, S=1/2 and three
nucleon resonances with L=1, S=3/2. Again, no evidence for spin-orbit
interactions.
We conclude that spin-orbit splittings are very weak and play no
decisive role for masses of baryon resonances.

\subsection{Regge trajectories, II}
We now present Figures which differ from standard Regge trajectories 
in choosing the orbital angular momentum L instead of J. Since
spin-orbit forces are small, the orbital angular momentum is well
defined. We choose L as variable since this allows us to combine
baryons of positive and negative parity.
\par
\begin{figure}[h]
\vspace*{-10mm}
\bc
\epsfig{file=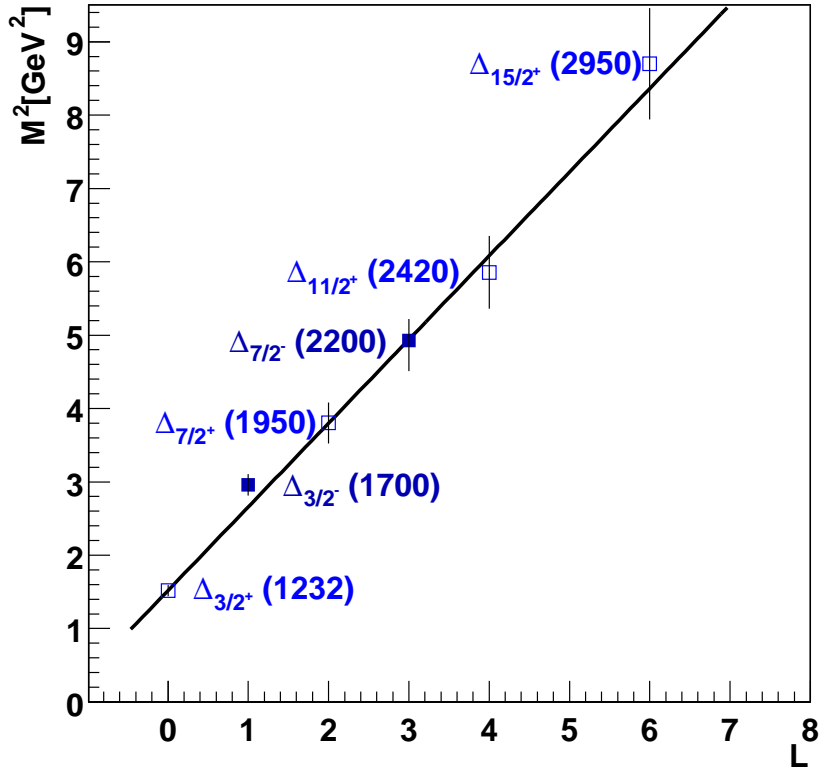,width=12cm}
\ec
\vspace*{-10mm}
\caption{$\Delta^*$'s with odd L and J=L+1/2 
fall on the same trajectory. \label{Delta-odd}}
\end{figure}

In Fig. \ref{Delta-odd} we include the  $\Delta_{3/2^-}$(1700) and the
$\Delta_{7/2^-}$(2220) to which we assign L=1, S=1/2 and
L=3, S=1/2, respectively. The two resonances have the lowest mass
with these quantum numbers. Their masses are fully compatible with the
Regge trajectory even though the intrinsic spin of the two resonances
with odd angular momentum is 1/2 whereas the other resonances have 
intrinsic spin 3/2. The spin-spin interaction within the mass
spectrum of $\Delta$ excitations thus vanishes or is small. Since
spin-orbit effects are small, we could have added the
$\Delta_{1/2^-}$(1620) at L=1. 

\par
\begin{figure}[h]
\vspace*{-10mm}
\bc
\epsfig{file=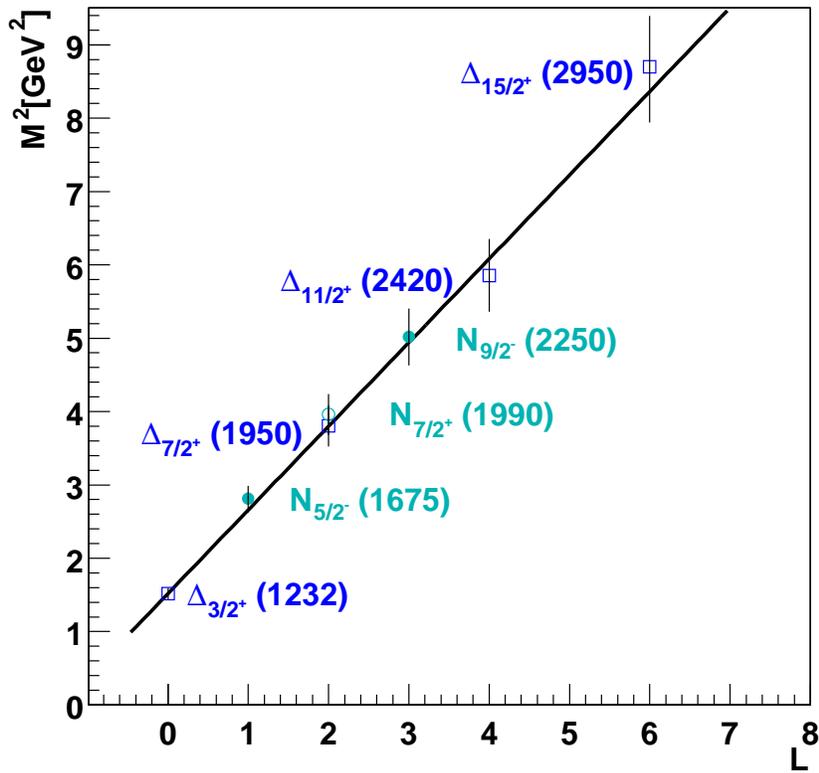,width=12cm}
\ec
\vspace*{-10mm}
\caption{N$^*$'s with intrinsic spin 3/2 fall on the same
trajectory.
\label{N-Delta}}
\end{figure}

Next we turn to nucleon resonances. In Fig. \ref{N-Delta} we have
included the nucleon resonances N$_{5/2^-}$(1675), N$_{7/2^+}$(1990),
and N$_{9/2^-}$(2220).  The N$_{5/2^-}$(1675),
N$_{3/2^-}$(1700), and N$_{1/2^-}$(1650) have similar masses and form
a super-multiplet with L=1, S=3/2. The N$_{5/2^-}$(1675) could also have 
L=3, and S=1/2 or S=3/2, the N$_{3/2^-}$(1700) L=3, S=3/2; but then
at least a N$_{7/2^-}$ in this mass range would be missing. A nucleon
resonance with these quantum numbers is observed at 2190 MeV; it is
accompanied by its own N$_{5/2^-}$ state (at 2200 MeV). 
A mixture of L=1 and L=3 both contributing to the wave function can of
course not be excluded but we assume L=1 to represent the
dominant part. 
\par
As seen in Fig. \ref{ls-split},
the N$_{7/2^+}$(1990) could be part of a super-multiplet with L=2 and S=3/2. 
The super-multiplet is approximately degenerate in mass with the
$\Delta$ super multiplet, also having  L=2 and S=3/2. 
The N$_{9/2^-}$(2220) also falls onto the general Regge trajectory
when we assume L=3, S=3/2 coupling to J=9/2. 
N and $\Delta$ resonances with a given L and with minimum mass fall on one
common Regge trajectory, except nucleon resonances with intrinsic spin 1/2. 
\par

\subsection{N$^*$ resonances with intrinsic spin S=1/2}
\begin{figure}[h]
\vspace*{-10mm}
\bc
\epsfig{file=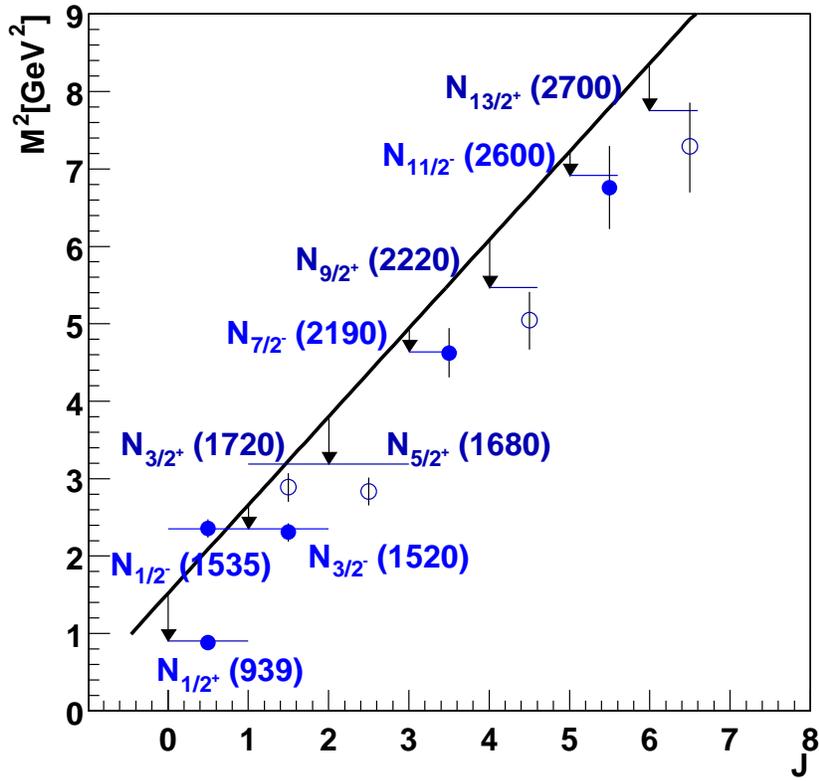,width=12cm}
\ec
\vspace*{-10mm}
\caption{The N$^*$ masses (with intrinsic spin S=1/2) 
lie below the standard Regge trajectory. They are
smaller by about 0.6 GeV$^2$ for N$^*$ in the 56-plet, and 
by 0.3 GeV$^2$ for N$^*$ in the 70-plet. \label{nucleons}}
\end{figure}

So far we have not considered the nucleon and nucleon resonances with
spin 1/2. In Fig. \ref{nucleons} we compare the squared masses of
positive- and negative-parity nucleon resonances to our standard Regge
trajectory. All resonances are lower in mass compared to the
trajectory. The mass shifts are visualized by arrows with a length
defined by the $\Delta (1232)$-N mass splitting; for nucleons with odd
angular momentum, the length of the arrow is divided by 2. The
vertical lines represent the expected masses, deduced from the
trajectory and a squared-mass shift calculated from 
\begin{equation}
\rm s_{i} = M^2_{\Delta (1232)} - M^2_{nucleon}.
\label{inst}
\end{equation}
We note that nucleons with S=1/2 are shifted in mass, nucleons with
spin 3/2 not. $\Delta$ excitations have not this spin-dependent mass
shift. The mass shift occurs only for baryons having spin and flavor
wave-functions 
which are both antisymmetric w.r.t. the exchange of two quarks. This is
the selection rule for instanton interactions which act only between
pairs of quarks which are antisymmetric w.r.t. their exchange in spin and
flavor \cite{Shuryak:bf}. 
We consider the even-odd staggering of Fig. \ref{inst} as
most striking evidence for the role of instanton interactions in
low-energy strong interactions.

\subsection{Octet-decuplet splitting}
\begin{figure}[h]
\vspace*{-10mm}
\bc
\epsfig{file=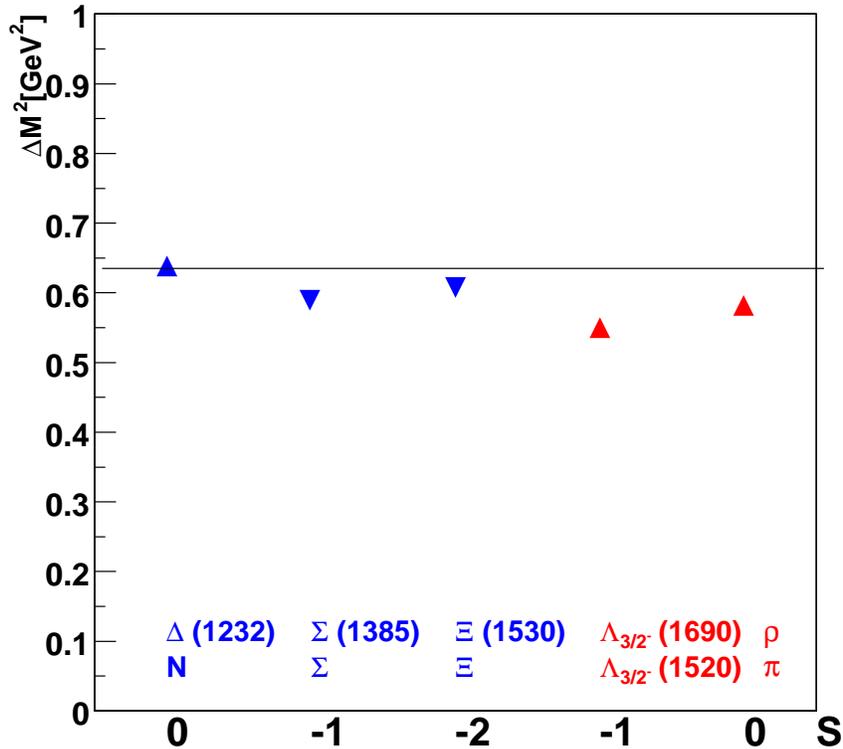,width=12cm}
\ec
\vspace*{-10mm}
\caption{The difference in squared masses of octet and decuplet
baryons. The difference is also shown for the two states
$\Lambda_{3/2^-}(1690)-\Lambda_{3/2^-}(1520)$ where the first one
belongs to the SU(3) octet, the latter to the singlet. The 
difference in mass square between $\rho$ and $\pi$ is of the same
order of magnitude: in all cases, the same forces act.
\label{oct-dec}}
\end{figure}

The N-$\Delta$ splitting as given in eq.(\ref{inst}) refers not only to
baryons composed of u and d quarks only. Fig. \ref{oct-dec} compares the
difference in mass square for $\Delta$-N, $\Sigma (1385)-\Sigma
(1195)$, and $\Xi (1530)-\Xi(1320)$. The three splittings are fully
compatible and evidence the flavor-independence of the strong
forces. Included are the mass square
differences between the octet and singlet
$\Lambda_{3/2^-}(1690)-\Lambda_{3/2^-}(1520)$ and of the $\rho-\pi$
system. Both $\Lambda$ resonances have masses which are influenced by
instanton-induced interactions. The $\Lambda_{3/2^-}(1690)$ is in a
70-plet where the antisymmetric component in the wave function is
reduced by a factor 1/2 compared to the nucleon. The
$\Lambda_{3/2^-}(1520)$ is a SU(6) singlet state, and antisymmetric in
all three quark-quark combinations. Hence this state is reduced in
the squared mass by 3/2 times the value given in (\ref{inst}). We may
thus expect, and find indeed, the same mass square difference as
observed for N and $\Delta$. We assign all these splittings to
instanton effects. 
\par
The  $\rho-\pi$ mass-square difference is shown to argue that the same
type of interaction is at work in mesons and baryons. This striking
similarity is not easily understood within the frame of present-day
models. It could indicate a deep symmetry between diquarks and quarks,
between bosons and fermions. Strong interactions remain invariant
when, in presence of a quark with spin up, a scalar diquark is
replaced by a quark with spin down, or when a vector diquark is
replaced by a quark with spin up \cite{Catto:wi}.
\clearpage
\subsection{Radial excitations}
Some partial waves show a second resonance at a higher mass. The best
known example is the Roper resonance, the N$_{1/2^+}$(1440). Its mass
is rather low compared to most calculations since, in the harmonic
oscillator description of baryon resonances, it is found in the second 
excitation band ($N=2$). 
\par
In meson spectroscopy, radial excitations do not have a mass shift
that is equivalent to the mass shift associated with two units of
orbital angular momentum. The $\rho_3(1690)$ has a much higher mass
than the $\rho (1450)$ (which is likely the radial excitation of the
$\rho (770)$, see however ref. \cite{Donnachie}). Based on a large
number of radial excitations, Bugg concluded 
\cite{Bugg} that the mean increase in
squared mass per radial excitation is $1.143\pm 0.009$\,GeV$^2$. This
is nearly the same value we determined from the mesonic Regge
trajectory. We assume that not only the slope of the Regge
trajectories of mesons and baryons are the same but also the spacings
between ground states and radially excited states. Thus
we determine the mass of the Roper resonance by adding to the squared
proton mass one unit of radial excitation energy and predict a Roper
mass of 1422 MeV. The second and third radial excitations are then
predicted to have masses of 1779 and 2076 MeV, respectively, to be
compared with  experimental candidates N$_{1/2^+}$(1710) and  
N$_{1/2^+}$(2100). Similarly, the first radial excitation of the 
$\Delta$(1232), $\Lambda (1115)$, $\Sigma (1193)$ and $\Xi (1320)$ are
predicted to have masses of 1631 (1600), 1565 (1600), 1565 (1560), and
1696 (1690) MeV; experimental values are quoted in parentheses.
\par
There is a band of negative-parity $\Delta$ resonances, 
$\Delta_{5/2^-}$(1930), $\Delta_{3/2^-}$(1940), $\Delta_{1/2^-}$(1900), 
with masses which correspond to the second excitation band. The
triplet of states is naturally interpreted as first radial excitation 
having intrinsic orbital angular momentum 1 and spin 3/2. They are
degenerate in mass with the two L=2 quartets 
N$_{1/2^+}(xxx)$, N$_{3/2^+}(1900)$, N$_{5/2^+}(2000)$,
N$_{7/2^+}(1990)$ and $\Delta_{1/2^+}(1910)$, $\Delta_{3/2^+}(1920)$,
$\Delta_{5/2^+}(1905)$, $\Delta_{7/2^+}(1950)$.
Again, one unit of orbital angular momentum gives the same excitation
energy as one unit in the radial quantum number. 
We note in passing that negative-parity $\Delta$ mesons with N=0 have
a total quark spin S=1/2; the symmetry of the SU(6) wave function
requires a wave 
function of mixed symmetry, i.e. the 70-plet. For N=1, the spatial
wave function can, even for L odd, be symmetric. The SU(6) wave
function can be symmetric and the $\Delta$ can be formed in a 56-plet
and thus have S=3/2.
\par  
Correspondingly, we assign the $\Delta_{9/2^-}(2400)$ to L=3, S=3/2,
N=1, and  the $\Delta_{13/2^-}(2750)$ to  L=5, S=3/2, N=1. The two 
states $\Delta_{7/2^-}(2200)$ and $\Delta_{5/2^-}(2350)$ should have
L=3. The latter is likely a partner of the $\Delta_{9/2^-}(2400)$,
the former could be the missing partner or
could have S=1/2, N=0 (which we assume).
\par
\begin{figure}[h]
\vspace*{-10mm}
\bc
\epsfig{file=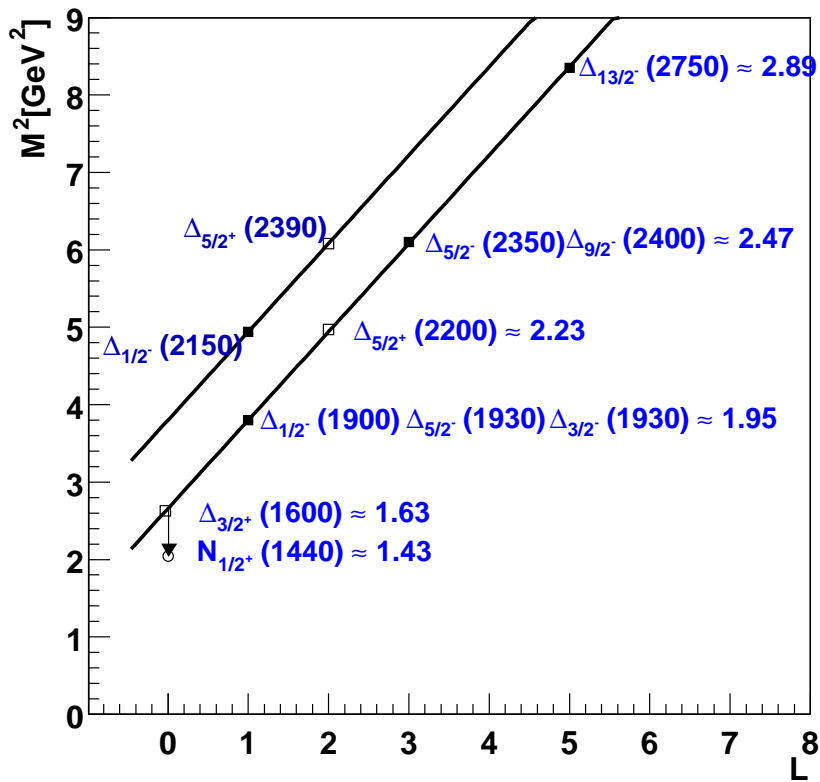,width=12cm}
\ec
\vspace*{-10mm}
\caption{The masses of $\Delta$ radial excitations are compatible with 
Regge trajectories shifted upwards by one unit [a] in mass-square
splitting. The symbols give the model-mass (also given numerically at
the right side). \label{radials}}
\end{figure}

 
Fig. \ref{radials} collects the leading radial excitations. For
clarity, the resonances are displayed as dots on the trajectory;
their experimental and expected masses are compared by giving
numbers. 
\subsection{Resonances with strangeness}
\begin{figure}[h]
\bc
\epsfig{file=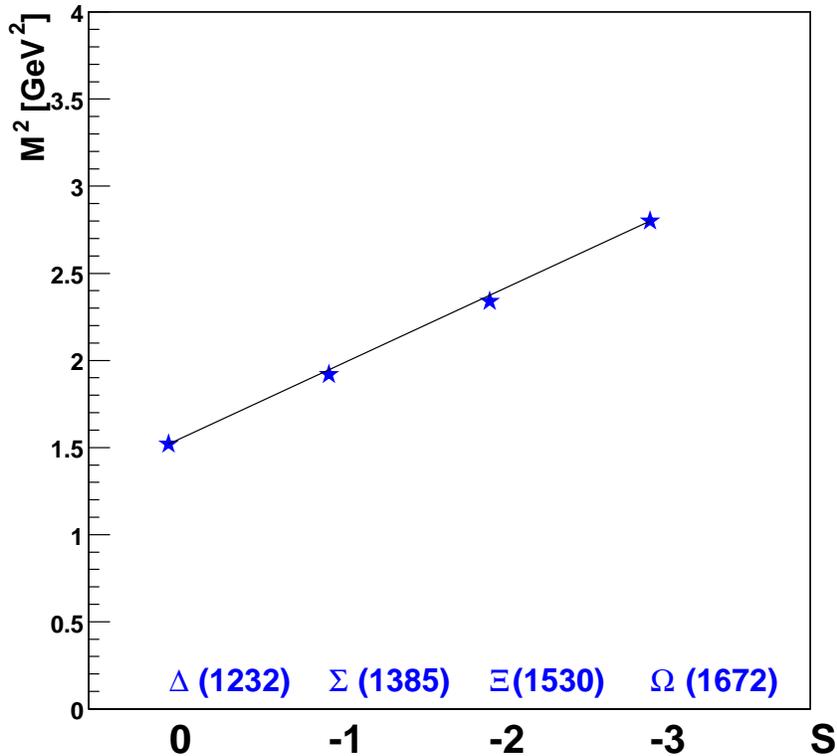,width=12cm}
\ec
\vspace*{-10mm}
\caption{The squared masses for the decuplet ground state baryons as a
function of their strangeness.
\label{strange}}
\end{figure}

The mass of a baryon increases with its strangeness content. This is
seen, e.g., in Fig. \ref{strange}. There are small deviations from the
linear interpolation when using squared masses, a linear mass
interpolation gives no better agreement. We use the quadratic form, 
as squared masses are linear in angular momentum and squared
masses are shifted by a constant value due to instanton interactions.

\vspace*{-10mm}
\vspace*{-5mm}

\subsection{Observations and conclusions}          
We now recall the basic experimental observations and draw obvious
conclusions from these facts. 
\begin{enumerate}
\item The slope of the Regge trajectory for meson- and
$\Delta$-excitations is identical.
Baryon resonances are quark-diquark excitations. 
\item $\ \Delta^*$ resonances with S=1/2 and S=3/2 are on
the same Regge trajectory.
There is no significant spin-spin splitting due to color-magnetic
interactions. Gluon exchange, often assumed to be responsible for the 
N-$\Delta$ splitting, should also lead to a mass shift of the 
$\Delta_{1/2^{-}}$(1620) and $\Delta_{3/2^{-}}$(1700) relative to the
Regge trajectory, in the same order of magnitude as in the case of the
N-$\Delta$ splitting. This is not the case. Gluon exchange is not
responsible for the N-$\Delta$ splitting.
\item N and $\Delta$ resonances with spin S=3/2 lie on
a common Regge trajectory. 
There is no genuine octet-decuplet splitting. 
For spin-3/2 resonances, there is no interaction
associated with the SU(6) multiplet structure. 
\item N$^*$'s and $\Delta^*$'s can be grouped into super-multiplets
with defined orbital angular momenta L and intrinsic spin S, but different
total angular momentum J. 
There is no significant spin-orbit ($\vec{\rm L}\cdot\vec{\rm S}$)
interaction. This is again an argument against a large role of gluon
exchange forces (even though the spin-orbit splitting due to one-gluon
exchange could be compensated by the Thomas precession in the
confinement potential). 
\item  Octet baryons with intrinsic spin 1/2 have a shift in the
squared mass. The shift is larger (by a factor 2) for even orbital
angular momenta than for odd angular momenta.
Wave functions of octet baryons with spin 1/2 contain
a component $\rm\left(q_1q_2 - q_2q_1\right)\left(\uparrow\downarrow\
- \downarrow\uparrow\right)$. 
The mass shift is proportional to this component. Instanton
interactions act on quark pairs which are antisymmetric in their spin
and their flavor wave function with respect to their exchange. The even-odd
mass shift visible in Fig. \ref{nucleons} manifests the importance of
instanton interactions in the baryon spectrum. 
\item Daughter trajectories have the same slope as the main trajectory
and an intercept which is higher by  $a=1.142$\,GeV$^2$ per $\ n$, both for
mesons and baryons. The similarity of the spacings between radial
excitations of mesons and baryons supports again the interpretation of 
baryon resonances as quark-diquark excitations. 
\end{enumerate}
\section{A new mass formula}
\label{sec_massformula}

\subsection{Introduction}
The observation that the mass of a baryon resonance is mostly given by
its internal orbital angular momentum and the fact that spin-spin and
spin-orbit splittings are small, except for instanton induced
interactions, allows us to write down a simple formula which
reproduces nearly all masses of baryon resonances observed so far. 
\par
\subsection{The mass formula}
The mass formula reads 
\vspace*{-8mm}
\begin{center}
\begin{equation}
\label{mass}
\rm  M^2 = M_{\Delta}^2 + \frac{n_s}{3}\cdot M_s^2  + 
a \cdot (L + N) - s_i \cdot I_{sym},
\end{equation}
\end{center}
where 
$$\rm M_s^2 = \left(M_{\Omega}^2 - M_{\Delta}^2\right), \qquad\ 
\rm s_i = \left(M_{\Delta}^2 - M_{\rm N}^2\right), $$
\noindent
$n_s$ the number of strange quarks in a baryon, and L the intrinsic orbital
angular momentum. N is the principal quantum number (we start with
N=0 for the ground state); L+2N gives
the harmonic-oscillator band {\em N}. $\rm I_{sym}$ is the fraction of the
wave function (normalized to the nucleon wave function) which is 
antisymmetric in spin and flavor. It is given by

\renewcommand{\arraystretch}{1.3}
\begin{tabular}{lrcc}
$\rm I_{sym}  = $&1.0& for S=1/2 and & octet in 56-plet; \\
$\rm I_{sym}  = $&0.5& for S=1/2 and & octet in 70-plet; \\
$\rm I_{sym}  = $&1.5& for S=1/2 and & singlet; \\
$\rm I_{sym}  = $&  0& otherwise.
\end{tabular}
\renewcommand{\arraystretch}{1.3}

$\rm M_{N}, M_{\Delta}, M_{\Omega}$ are input parameters taking from PDG,
$a = 1.142$/GeV$^2$ is the Regge slope as determined from the meson spectrum.
\par
For a quantitative comparison between data and the mass formula,
masses and errors need to be defined. The Particle Data Group lists ranges of
acceptable values; we use the central value for the comparison. Our
error consists of two parts, of one model error of 30 MeV and one
width-dependent error. The model error avoids extremely large $\chi^2$
contributions from the octet ground-state particles. The second error
allows for mass shifts of resonances due to hadronic effects, like
virtual decays or couplings to close-by thresholds. 
We estimate this effect to be of the order of one
quarter of the width, and use $\Gamma$/4 as second error
contribution even though we know that strong couplings to two-particle
thresholds may result in much larger mass shifts. 
The two errors are added quadratically. 
\par
The widths of the resonances are often not well determined and, for
less established baryons, no width estimate is given by the Particle
Data Group. We parameterize all widths using the formula
\vspace*{-8mm}
\begin{center}
\begin{equation}
\label{width}
\rm \Gamma = \frac{Q}{4} 
\end{equation}
\end{center}
where $\rm Q$ is the largest available energy for hadronic decays.
 
\subsection{Comparison with data}
In the following Tables, we give a quantitative comparison between data
and the mass formula. Overall we find $\chi^2$=117 for 97 degrees of
freedom. In total, there are 103 entries in the Tables below. Three
data are used to define the model: the masses of N, $\Delta$ and
$\Omega$. Three states are not reproduced by the model: 
the $\Sigma (1480)$ with 1*, the $\Xi (1620)$ with 1*, and the 
$\Omega (2380)$ with 2*'s. None of them has known spin-parity. They
were observed as bumps in invariant mass spectra. 
\par
First we recognize that the overall consistency of data and model is
excellent. There is no free parameter used in describing the data:
three baryon masses are used as input values, the slope parameter a 
of the Regge trajectories is taken from the meson spectrum. The
$\chi^2$ achieved depends of course very critically on the error
choice. A constant error contribution is needed to get the $\Lambda$
and $\Sigma$ to be compatible with one common value; there is no
parameter to describe their splitting. Such a parameter would have a
bad effect on high-mass states: the two 
$\Lambda_{1/2^-}(1800)$, $\Lambda_{5/2^-}(1830)$ have a higher mass
than the triplet $\Sigma_{1/2^-}(1750)$, $\Sigma_{3/2^-}(1670)$,
$\Sigma_{3/2^-}(1775)$.  
\par
Now we discuss discrepancies beyond 2$\sigma$, $\chi^2 > 4$. The 
$\Delta_{1/2^+}(1750)$ and $\Sigma_{7/2^-}(2100)$ are 1* resonances,
and the discrepancy does not need to be a failure of the model. 
For the $\Lambda_{1/2^+}(1600)$ and $\Lambda_{1/2^+}(1810)$, both 3*
resonances, the 
errors given by $\Gamma$/4 are underestimated; the mass of the 
$\Lambda_{1/2^+}(1600)$ falls into the range from 1560 to 1700 MeV,
the predicted mass is 1565 MeV. The $\Lambda_{1/2^+}(1810)$ should
have a mass in the 1750 to 1850 range; the predicted value of 1895 MeV
is still larger but now compatible within the model error. 
The two resonances $\Xi(2250)$ and $\Omega(2380)$ have no known
spin-parities; it is therefore difficult to appreciate the meaning of
the discrepancy. It is not excluded that in baryon resonances with two or
three strange quarks, heavy-quark physics is starting to take over,
that gluon exchange begins to be effective and that the extrapolation
of the mass formula to $\Xi$ and $\Omega$ states is not
justified. Clearly, there is not sufficient experimental information
to clarify this point in a phenomenological description of data.
\par
The $\Sigma_{3/2^-}(1580)$ and $\Sigma_{3/2^-}(1670)$ are more
critical. The $\Sigma_{3/2^-}(1670)$ is a 4* resonance with
a well-measured mass. It would perfectly fit, with the
$\Sigma_{1/2^-}(1620)$, as $(70,^28)_1$ instead of  $(70,^48)_1$
resonance. But then, the 2* state $\Sigma_{3/2^-}(1580)$ would have no
slot. If we remove it, the total $\chi^2$ contribution would go down
from 34.69 to 23.05 (for now 24 degrees of freedom). A 2* resonance
should perhaps not be 'talked away'. But the experimental situation is
certainly not clear enough to reject the model because of these two
$\Sigma$ states. 

\begin{table}
\caption{Mass spectrum of N resonances. 
A nucleon resonances is characterized by it $\rm J^P$ as subscript 
and its nominal mass (in parenthesis). The PDG rating is given by the
number of *'s. Its classification into multiplets is discussed in
section (6). The PDG lists a range of acceptable
values, we give the central mass (in MeV), compared to the predicted
mass from eq. (\protect\ref{mass}). We list the PDG range of acceptable
widths $\Gamma$ and compare them to eq. (\protect\ref{width}). The width
parameterization is only used to estimate errors. The mass errors $\sigma$
are given by $\sigma^2 = \frac{\Gamma^2}{16} + 30^2$ where the
first error allows for hadronic mass shifts in the order of 1/4 of
the line width, the second one for uncertainties in the mass formula.
The last column gives the $\chi^2$ contribution from the mass
comparison. The $\chi^2$'s are summed up and compared to the degrees
of freedom in the last column.
}
\bc
\renewcommand{\arraystretch}{1.3}
\begin{tabular}{||cccc|cccccc||}
\hline
\hline
Baryon            &Status& $\rm D_L$  & N & Mass  &(\ref{mass}) &
$\Gamma$ & (\ref{width}) &$\sigma$ & $\chi^2$ \\ \hline
N$_{1/2^+}(939)$ & **** & $(56,^28)_0$ & 0 & 939 & - & - & - & - & - \\
N$_{1/2^+}(1440)$ & **** & $(56,^28)_0$ & 1 & 1450 & 1423 & 250-450 & 87 & 37 & 0.53\\
N$_{1/2^+}(1710)$ & *** & $(56,^28)_0$ & 2 & 1710 & 1779 & 50-250 & 176 & 53 & 1.69\\
$^1$N$_{1/2^+}(2100)$ & * & $(56,^28)_0$ & 2 & 2100 & 2076 & - & 251 & 70 & 0.12\\ \hline
N$_{1/2^-}(1535)$ & **** & $(70,^28)_1$ & 0 & 1538 & 1530 & 100-250 & 114 & 41 & 0.04\\
N$_{3/2^-}(1520)$ & **** & $(70,^28)_1$ & 0 & 1523 & 1530 & 110-135 & 114 & 41 & 0.03\\ \hline
N$_{1/2^-}(1650)$ & **** & $(70,^48)_1$ & 0 & 1660 & 1631 & 145-190 & 139 & 46 & 0.4\\
N$_{3/2^-}(1700)$ & *** & $(70,^48)_1$ & 0 & 1700 & 1631 & 50-150 & 139 & 46 & 2.25\\
N$_{5/2^-}(1675)$ & **** & $(70,^48)_1$ & 0 & 1678 & 1631 & 140-180 & 139 & 46 & 1.04\\ \hline
N$_{3/2^+}(1720)$ & **** & $(56,^28)_2$ & 0 & 1700 & 1779 & 100-200 & 176 & 53 & 2.22\\
N$_{5/2^+}(1680)$ & **** & $(56,^28)_2$ & 0 & 1683 & 1779 & 120-140 & 176 & 53 & 3.28\\ \hline
N$_{3/2^+}(1900)$ & ** & $(70,^48)_2$ & 0 & 1900 & 1950 & - & 219 & 62 & 0.65\\
N$_{5/2^+}(2000)$ & ** & $(70,^48)_2$ & 0 & 2000 & 1950 & - & 219 & 62 & 0.65\\
N$_{7/2^+}(1990)$ & ** & $(70,^48)_2$ & 0 & 1990 & 1950 & - & 219 & 62 & 0.42\\ \hline
N$_{1/2^-}(2090)$ & * & $(70,^28)_1$ & 2 & 2090 & 2151 & - & 269 & 74 & 0.68\\
N$_{3/2^-}(2080)$ & ** & $(70,^28)_1$ & 2 & 2080 & 2151 & - & 269 & 74 & 0.92\\ \hline 
N$_{5/2^-}(2200)$ & ** & $(70,^28)_3$ & 0 & 2220 & 2151 & - & 269 & 74 & 0.87\\ 
N$_{7/2^-}(2190)$ & **** & $(70,^28)_3$ & 0 & 2150 & 2151 & 350-550 & 269 & 74 & 0\\ \hline 
N$_{9/2^-}(2250)$ & **** & $(70,^48)_3$ & 0 & 2240 & 2223 & 290-470 & 287 & 78 & 0.05\\ \hline 
N$_{9/2^+}(2220)$ & **** & $(56,^28)_4$ & 0 & 2245 & 2334 & 320-550 & 315 & 84 & 1.12\\ \hline
N$_{11/2^-}(2600)$ & *** & $(70,^28)_5$ & 0 & 2650 & 2629 & 500-800 & 389 & 102 & 0.04\\ \hline
N$_{13/2^+}(2700)$ & ** & $(56,^28)_6$ & 0 & 2700 & 2781 & - & 427 & 111 & 0.53\\ \hline \hline
& & & & & & dof: & 21 & $\sum \chi^2$:& 17.53 \\
\hline
\hline
\end{tabular}
\renewcommand{\arraystretch}{1.0}
\ec
$^1$\small{
Based on its mass, the N$_{1/2^+}(2100)$ is likely a radial
excitation. It could also be the $(70,^48)_2$ N$_{1/2^+}$ state
expected at 1950 MeV. The SAPHIR collaboration suggested a  N$_{1/2^+}$
at 1986 MeV \protect\cite{Plotzke:ua} 
which would, if confirmed, be a natural
partner to complete the quartet of L=2, S=3/2 nucleon resonances.}
\label{n}
\end{table}

\begin{table}
\caption{Mass spectrum of $\Delta$ resonances. See caption of Table 
(\protect\ref{n}).}
\bc
\renewcommand{\arraystretch}{1.4}
\begin{tabular}{||cccc|cccccc||}
\hline
\hline
Baryon            &Status& $\rm D_L$  & N & Mass  &(\ref{mass}) &
$\Gamma$ & (\ref{width}) &$\sigma$ & $\chi^2$ \\ \hline
$\Delta_{3/2^+}(1232)$ & **** & $(56,^410)_0$ & 0 & 1232 & 1232 & - & - & - & -\\
$\Delta_{3/2^+}(1600)$ & *** & $(56,^410)_0$ & 1 & 1625 & 1631 & 250-450 & 139 & 46 & 0.02\\ \hline
$\Delta_{1/2^+}(1750)$ & * & $(70,^210)_0$ & 1 & 1750 & 1631 & - & 139 & 46 & 6.69\\ \hline
$\Delta_{1/2^-}(1620)$ & **** & $(70,^210)_1$ & 0 & 1645 & 1631 & 120-180 & 139 & 46 & 0.09\\
$\Delta_{3/2^-}(1700)$ & **** & $(70,^210)_1$ & 0 & 1720 & 1631 & 200-400 & 139 & 46 & 3.74\\ \hline
$\Delta_{1/2^-}(1900)$ & ** & $(56,^410)_1$ & 1 & 1900 & 1950 & 140-240 & 219 & 62 & 0.65\\
$\Delta_{3/2^-}(1940)$ & * & $(56,^410)_1$ & 1 & 1940 & 1950 & - & 219 & 62 & 0.03\\
$\Delta_{5/2^-}(1930)$ & *** & $(56,^410)_1$ & 1 & 1945 & 1950 & 250-450 & 219 & 62 & 0.01\\ \hline
$\Delta_{1/2^+}(1910)$ & **** & $(56,^410)_2$ & 0 & 1895 & 1950 & 190-270 & 219 & 62 & 0.79\\
$\Delta_{3/2^+}(1920)$ & *** & $(56,^410)_2$ & 0 & 1935 & 1950 & 150-300 & 219 & 62 & 0.06\\
$\Delta_{5/2^+}(1905)$ & **** & $(56,^410)_2$ & 0 & 1895 & 1950 & 280-440 & 219 & 62 & 0.79\\
$\Delta_{7/2^+}(1950)$ & **** & $(56,^410)_2$ & 0 & 1950 & 1950 & 290-350 & 219 & 62 & 0\\ \hline
$\Delta_{1/2^-}(2150)$ & * & $(70,^210)_1$ & 2 & 2150 & 2223 & - & 287 & 78 & 0.88\\ \hline
$\Delta_{7/2^-}(2200)$ & * & $(70,^210)_3$ & 0 & 2200 & 2223 & - & 287 & 78 & 0.09\\ \hline
$^1\Delta_{5/2^+}(2000)$ & ** & $(70,^210)_2$ & 1 & 2200 & 2223 & - & 287 & 78 & 0.09\\ \hline
$\Delta_{5/2^-}(2350)$ & * & $(56,^410)_3$ & 1 & 2350 & 2467 & - & 348 & 92 & 1.62\\
$\Delta_{9/2^-}(2400)$ & ** & $(56,^410)_3$ & 1 & 2400 & 2467 & - & 348 & 92 & 0.53\\ \hline
$\Delta_{7/2^+}(2390)$ & * & $(56,^410)_4$ & 0 & 2390 & 2467 & - & 348 & 92 & 0.7\\
$\Delta_{9/2^+}(2300)$ & ** & $(56,^410)_4$ & 0 & 2300 & 2467 & - & 348 & 92 & 3.3\\
$\Delta_{11/2^+}(2420)$ & **** & $(56,^410)_4$ & 0 & 2400 & 2467 & 300-500 & 348 & 92 & 0.53\\ \hline
$\Delta_{13/2^-}(2750)$ & ** & $(56,^410)_5$ & 1 & 2750 & 2893 & - & 455 & 118 & 1.47\\ \hline
$\Delta_{15/2^+}(2950)$ & ** & $(56,^410)_6$ & 0 & 2950 & 2893 & - & 455 & 118 & 0.23\\ \hline \hline
& & & & & & dof: & 21 & $\sum \chi^2$:& 22.31 \\
\hline
\hline
\end{tabular}
\renewcommand{\arraystretch}{1.0}
\ec
$^1$\small{The PDG quotes two entries, at 1752 and 2200 MeV, respectively,
and gives 2000 as ''our estimate''. 
We use the higher mass value for our comparison.}
\label{d}
\end{table}
\begin{table}
\caption{Mass spectrum of $\Lambda$ resonances. See caption of Table
(\protect\ref{n}).}
\bc
\renewcommand{\arraystretch}{1.4}
\begin{tabular}{||cccc|cccccc||}
\hline
\hline
Baryon            &Status& $\rm D_L$  & N & Mass  &(\ref{mass}) &
$\Gamma$ & (\ref{width}) &$\sigma$ & $\chi^2$ \\ \hline
$\Lambda_{1/2^+}(1115)$ & **** & $(56,^28)_0$ & 0 & 1116 & 1144 & - & - & 30 & 0.87\\
$\Lambda_{1/2^+}(1600)$ & *** & $(56,^28)_0$ & 1 & 1630 & 1565 & 50-250 & 32 & 31 & 4.4\\
$\Lambda_{1/2^+}(1810)$ & *** & $(56,^28)_0$ & 2 & 1800 & 1895 & 50-250 & 115 & 42 & 5.12\\ \hline
$\Lambda_{1/2^-}(1405)$ & **** & $(70,^21)_1$ & 0 & 1407 & 1460 & 50 & 6 & 30 & 3.12\\
$\Lambda_{3/2^-}(1520)$ & **** & $(70,^21)_1$ & 0 & 1520 & 1460 & 16 & 6 & 30 & 4\\ \hline
$\Lambda_{1/2^-}(1670)$ & **** & $(70,^28)_1$ & 0 & 1670 & 1664 & 25-50 & 57 & 33 & 0.03\\
$\Lambda_{3/2^-}(1690)$ & **** & $(70,^28)_1$ & 0 & 1690 & 1664 & 50-70 & 57 & 33 & 0.62\\ \hline
$\Lambda_{1/2^-}(1800)$ & *** & $(70,^48)_1$ & 0 & 1785 & 1757 & 200-400 & 80 & 36 & 0.6\\
$\Lambda_{5/2^-}(1830)$ & **** & $(70,^48)_1$ & 0 & 1820 & 1757 & 60-110 & 80 & 36 & 3.06\\ \hline
$\Lambda_{3/2^+}(1890)$ & **** & $(56,^28)_2$ & 0 & 1880 & 1895 & 60-200 & 115 & 42 & 0.13\\
$\Lambda_{5/2^+}(1820)$ & **** & $(56,^28)_2$ & 0 & 1820 & 1895 & 70-90 & 115 & 42 & 3.19\\ \hline
$\Lambda(2000)$ & * & $(70,^48)_2$ & 0 & 2000 & 2056 & - & 155 & 49 & 1.31\\
$\Lambda_{5/2^+}(2110)$ & *** & $(70,^48)_2$ & 0 & 2115 & 2056 & 150-250 & 155 & 49 & 1.45\\
$\Lambda_{7/2^+}(2020)$ & * & $(70,^48)_2$ & 0 & 2020 & 2056 & - & 155 & 49 & 0.54\\
\hline
$\Lambda_{7/2^-}(2100)$ & **** & $(70,^21)_3$ & 0 & 2100 & 2101 & 100-250 & 166 & 51 & 0\\ \hline
$\Lambda_{3/2^-}(2325)$ & * & $(70,^28)_1$ & 2 & 2325 & 2248 & - & 203 & 59 & 1.7\\ \hline
$\Lambda_{9/2^+}(2350)$ & *** & $(56,^28)_4$ & 0 & 2355 & 2424 & 100-250 & 247 & 69 & 1\\ \hline
$\Lambda(2585)$ & ** & $(70,^48)_2$ & 0 & 2585 & 2551 & - & 279 & 76 & 0.2\\ \hline \hline
& & & & & & dof: & 18 & $\sum \chi^2$:& 31.34 \\
\hline
\hline
\end{tabular}
\renewcommand{\arraystretch}{1.0}
\ec
\label{l}
\end{table}
\begin{table}
\caption{Mass spectrum of $\Sigma$ resonances. See caption of Table
(\protect\ref{n}).}
\bc
\renewcommand{\arraystretch}{1.4}
\begin{tabular}{||cccc|cccccc||}
\hline
\hline
Baryon            &Status& $\rm D_L$  & N & Mass  &(\ref{mass}) &
$\Gamma$ & (\ref{width}) &$\sigma$ & $\chi^2$ \\ \hline
$\Sigma_{1/2^+}(1193)$ & **** & $(56,^28)_0$ & 0 & 1193 & 1144 & - & - & 30 & 2.67\\
$\Sigma_{3/2^+}(1385)$ & **** & $(56,^410)_0$ & 0 & 1384 & 1394 & - & - & 30 & 0.11\\
$\Sigma(1480)$ & * &  &  &  &  &  &  &  & \\
$\Sigma(1560)$ & ** & $(56,^28)_0$ & 1 & 1560 & 1565 & - & 32 & 31 & 0.03\\
$\Sigma_{1/2^+}(1660)$ & *** & $(70,^28)_0$ & 1 & 1660 & 1664 & 40-200 & 57 & 33 & 0.01\\
$\Sigma_{1/2^+}(1770)$ & * & $(70,^210)_0$ & 1 & 1770 & 1757 & - & 80 & 36 & 0.13\\
$\Sigma_{1/2^+}(1880)$ & ** & $(56,^28)_0$ & 2 & 1880 & 1895 & - & 115 & 42 & 0.13\\ \hline
$\Sigma_{1/2^-}(1620)$ & ** & $(70,^28)_1$ & 0 & 1620 & 1664 & - & 57 & 33 & 1.78\\
$\Sigma_{3/2^-}(1580)$ & ** & $(70,^28)_1$ & 0 & 1580 & 1664 & - & 57 & 33 & 6.48\\
$\Sigma(1690)$ & ** & $(70,^210)_1$ & 0 & 1690 & 1757 & - & 80 & 36 & 3.46\\ \hline
$\Sigma_{1/2^-}(1750)$ & *** & $(70,^48)_1$ & 0 & 1765 & 1757 & 60-160 & 80 & 36 & 0.05\\
$\Sigma_{3/2^-}(1670)$ & **** & $(70,^48)_1$ & 0 & 1675 & 1757 & 40-80 & 80 & 36 & 5.19\\
$\Sigma_{5/2^-}(1775)$ & **** & $(70,^48)_1$ & 0 & 1775 & 1757 & 105-135 & 80 & 36 & 0.25\\ \hline
$\Sigma_{1/2^-}(2000)$ & * & $(70,^28)_1$ & 1 & 2000 & 1977 & - & 135 & 45 & 0.26\\
$\Sigma_{3/2^-}(1940)$ & *** & $(70,^28)_1$ & 1 & 1925 & 1977 & 150-300 & 135 & 45 & 1.34\\ \hline
$\Sigma_{3/2^+}(1840)$ & * & $(56,^28)_2$ & 0 & 1840 & 1895 & - & 115 & 42 & 1.71\\
$\Sigma_{5/2^+}(1915)$ & **** & $(56,^28)_2$ & 0 & 1918 & 1895 & 80-160 & 115 & 42 & 0.3\\ 
\hline
$^{1}\Sigma_{3/2^+}(2080)$ & ** & $(56,^410)_2$ & 0 & 2080 & 2056 & - & 155 & 49 & 0.24\\
$^{1}\Sigma_{5/2^+}(2070)$ & * & $(56,^410)_2$ & 0 & 2070 & 2056 & - & 155 & 49 & 0.06\\
$^{1}\Sigma_{7/2^+}(2030)$ & **** & $(56,^410)_2$ & 0 & 2033 & 2056 & 150-200 & 155 & 49 & 0.22\\ 
\hline
$\Sigma(2250)$ & *** & $(70,^28)_3$ & 0 & 2245 & 2248 & 60-150 & 203 & 59 & 0\\
$\Sigma_{7/2^-}(2100)$ & * & $(70,^28)_3$ & 0 & 2100 & 2248 & - & 203 & 59 & 6.29\\ \hline
$\Sigma(2455)$ & ** & $(56,^28)_4$ & 0 & 2455 & 2424 & - & 247 & 69 & 0.2\\ \hline
$\Sigma(2620)$ & ** & $(70,^28)_5$ & 0 & 2620 & 2708 & - & 318 & 85 & 1.07\\ \hline
$\Sigma(3000)$ & * & $(56,^28)_6$ & 0 & 3000 & 2857 & - & 355 & 94 & 2.31\\ \hline
$\Sigma(3170)$ & * & $(70,^28)_7$ & 0 & 3170 & 3102 & - & 416 & 108 & 0.4\\ \hline \hline
& & & & & & dof: & 25 & $\sum \chi^2$:& 34.69 \\ 
\hline
\hline
\end{tabular}
\renewcommand{\arraystretch}{1.0}
\ec
$^1$\small{ These three resonances, and the missing $\Sigma_{1/2^+}$, can
belong to the octet or to the decuplet; the mass formula
(\protect\ref{mass}) predicts identical masses.}
\label{s}
\end{table}
\begin{table}
\caption{Mass spectrum of $\Xi$ resonances. See caption of Table
(\protect\ref{n}).}
\bc
\renewcommand{\arraystretch}{1.4}
\begin{tabular}{||cccc|cccccc||}
\hline
\hline
Baryon            &Status& $\rm D_L$  & N & Mass  &(\ref{mass}) &
$\Gamma$ & (\ref{width}) &$\sigma$ & $\chi^2$ \\ \hline
$\Xi_{1/2^+}(1320)$ & **** & $(56,^28)_0$ & 0 & 1315 & 1317 & - & - & 30 & 0\\
$\Xi_{3/2^+}(1530)$ & **** & $(56,^410)_0$ & 0 & 1532 & 1540 & 9 & - & 30 & 0.07\\
$\Xi(1620)$ & * &  &  & 1620 &  & &  &  & \\
$\Xi(1690)$ & *** & $(56,^28)_0$ & 1 & 1690 & 1696 & $<$30 & 21 & 30 & 0.04\\ \hline
$\Xi_{3/2^-}(1820)$ & *** & $(70,^28)_1$ & 0 & 1823 & 1787 & 14-39 & 43 & 32 & 1.27\\ \hline
$\Xi(1950)$ & *** & $(56,^28)_2$ & 0 & 1950 & 2004 & 40-80 & 98 & 39 & 1.92\\ 
$\Xi(2030)$ & *** & $(56,^28)_2$ & 0 & 2025 & 2004 & 15-35 & 98 & 39 & 0.29\\ \hline
$\Xi(2120)$ & * & $(56,^410)_2$ & 0 & 2120 & 2157 & - & 136 & 45 & 0.68\\ 
$\Xi(2250)$ & ** & $(56,^410)_2$ & 0 & 2250 & 2157 & - & 136 & 45 & 4.27\\ \hline
$\Xi(2370)$ & ** & $(70,^28)_3$ & 0 & 2370 & 2340 & - & 182 & 55 & 0.3\\ 
$\Xi(2500)$ & * & $(56,^28)_4$ & 0 & 2500 & 2510 & - & 224 & 64 & 0.02\\ \hline \hline
& & & & & & dof: & 10 & $\sum \chi^2$:& 8.86 \\ 
\hline
\hline
\end{tabular}
\renewcommand{\arraystretch}{1.0}
\ec
\label{x}
\end{table}
\begin{table}
\caption{Mass spectrum of $\Omega$ resonances. See caption of Table
(\protect\ref{n}).}
\bc
\renewcommand{\arraystretch}{1.4}
\begin{tabular}{||cccc|cccccc||}
\hline
\hline
Baryon            &Status& $\rm D_L$  & N & Mass  &(\ref{mass}) &
$\Gamma$ & (\ref{width}) &$\sigma$ & $\chi^2$ \\ \hline
$\Omega_{3/2^+}(1672)$ & **** & $(56,^410)_0$ & 0 & 1672 & - & - & - & - & - \\
$\Omega(2250)$ & **** & $(56,^410)_2$ & 0 & 2252 & 2254 & 37-73 & 77 & 36 & 0\\
$\Omega(2380)$ & ** & - & - & 2380 & - & - & - & - & -\\
$\Omega(2470)$ & ** & $(70,^210)_3$ & 0 & 2474 & 2495 & 39-105 & 137 & 46 & 0.21\\ \hline \hline
& & & & & & dof: & 2 & $\sum \chi^2$:& 0.21 \\
\hline
\hline
\end{tabular}
\renewcommand{\arraystretch}{1.0}
\ec
\label{o}
\end{table}

\clearpage
\section{Multiplet structure of observed and missing resonances}
\subsection{Missing resonances}
We now discuss how baryon resonances can be assigned to given
multiplets. The emphasis of this discussion will be to identify the
nature of the so-called $missing$ resonances. There are different
reasons for resonances not to be observed, and theoretical guidance
(or prejudices) are needed to understand why a particular resonance
has not been observed. We distinguish between different classes of
$missing$ resonances:
\begin{enumerate}
\item {\em Trivially missing resonances}, resonances which are expected to
exist in a model-independent way. E.g., there is no known $\Omega_{3/2^-}$
state even though nobody will doubt that it would be discovered in an
appropriate experiment.
\item There are {\em hidden} resonances, resonances with identical
quantum numbers having, according to our mass formula (\ref{mass}),
the same mass but differing in their internal spin-flavor structure. 
From the nucleon resonances at 1950
MeV we know that there is a $^48$ multiplet at about 1950 MeV. They
belong to a 70-plet. We must therefore expect a  $\Delta$
spin doublet $^210$. According to (\ref{mass}) the $^210$ doublet has
the same mass as the spin quartet $^410$. There are hence two 
$\Delta_{3/2^+}$(1950) and two $\Delta_{5/2^+}$(1950) states 
expected. The second radial excitation of the $\Delta (1232)$ is also 
expected as $\Delta_{3/2^+}$(1950). A careful high-statistics study of 
several decay modes could possibly reveal that more than one resonance
contributes; at the present level of experiments they are unobservable.
We call these missing resonances {\em hidden} resonances. 
\item Above 2.5 GeV, only stretched resonances are observed, with spin
and orbital angular momentum parallel. Nucleon resonances have spin
1/2 and J=L+1/2; $\Delta$ excitations prefer S=3/2 and J=L+3/2. Since
S=3/2 is forbidden for N=0, $\Delta$'s with S=3/2 must have one unit 
of radial
excitation. Clearly, other ($\vec{\rm L}\cdot\vec{\rm S}$) couplings are
possible and two nucleon- or four $\Delta$-resonances with
approximately the same mass but different J could exist. These states
are solutions of the Hamiltonian; but this does not guarantee that they
are realized dynamically as stable rotations. 
A match box has three axes of rotation,
dynamically realized are only those around the axis of minimal and
maximal moment of inertia.  We call resonances expected as solutions
of the Hamiltonian but not realized dynamically {\em missing
resonances}. Of course, {\em hidden} resonances can also be suppressed
dynamically and thus belong to the class of {\em missing resonances}.
\end{enumerate}
\subsection{\label{groundstates}Ground states}
The ground states of octet and decuplet baryons are, of course, all
experimentally well established. The N, $\Delta (1232)$, and $\Omega$
masses are used as input parameters. There is no parameter allowing
for a $\Lambda (1115)-\Sigma (1193)$ splitting which amounts to 77
MeV. The model predicts 1144 MeV. For resonances, there is no general
mass enhancement of $\Sigma^*$'s compared to $\Lambda^*$'s. The
$\Sigma (1385)$ and $\Xi (1530)$ decuplet ground states are rather
well reproduced in the model.  
\subsection{\label{scalar}Radial excitations of the ground states}
\par
We have seen that there is a sequence of spin 1/2$^+$ nucleon
and spin 3/2$^+$ $\Delta$ resonances which can be assigned to radial
excitations of the respective ground state. These, and corresponding
states for the hyperons, are collected in Table \ref{tab:radial}.
In the
harmonic-oscillator representation of radially excited states, there
are two types of resonances expected in the second band, having
N$_{1/2^+}$ quantum numbers with $\rm (D,L_N^P) = (56,0_{2}^{+})$ and
$(70,0_{2}^{+})$ SU(6) wave functions. Decuplet baryons in the 56- and
70-plet are expected to be mass-degenerate; octet baryons in 56 or 70
feel different instanton-induced interactions. The members of the
70-plet would be N$_{1/2^+}(1530)$, $\Lambda_{1/2^+}(1660)$,
$\Sigma_{1/2^+}(1660)$ and $\Xi_{1/2^+}(1787)$. In the decuplet we
expect $\Delta_{1/2^+}(1631)$, $\Delta_{1/2^+}(1950)$, and a
$\Sigma_{1/2^+}(1757)$. One radially excited $\Xi$ state is observed,
further states and $\Omega$ resonances are {\em trivially} 
missing. 
\par
We note that N and $\Delta$ radial excitations are mostly compatible 
with an assignment to a 56-plet and not to 70-plets. The 1* 
$\Delta_{1/2^+}(1750)$ is an exception of this rule. It is not clear 
if there are really two resonances $\Sigma_{1/2^+}(1660)$ and 
$\Sigma_{1/2^+}(1770)$ but both need to be assigned to a radial
excitation in a 70-plet. Hence we believe that the 70-plet is needed
to complete the spectrum of radially excited states. The absence of 
radially excited nucleon states in the 70-plet could reflect a
dynamical suppression: the 70-plet could possibly
be formed only in the  case of
unequal quark masses. This conjecture would require the
$\Delta_{1/2^+}(1750)$ not to exist and remove the largest
single $\chi^2$ contribution. This question 
certainly needs theoretical study. 
For the ground states, the 70-plet is of course forbidden due to
symmetry reasons. 
\par

\begin{table}[h!]
\caption{SU(3) octet and decuplet radial excitations of baryon resonances.  
The 6$^{\rm th}$ and the last column give masses 
as calculated from eq. (\protect\ref{mass}).
\label{tab:radial}}
\renewcommand{\arraystretch}{1.4}
\bc
\begin{tabular}{|lcc|cc|c|cc|c|} 
\hline
L=0&N=0 & 56  & $^28$  & N$_{1/2^+}(939)$   &939 & $^410$ & $\Delta_{3/2^+}(1232)$ &1232\\ 
L=0&N=1 & 56  & $^28$  & N$_{1/2^+}(1440)$  &1422& $^410$ & $\Delta_{3/2^+}(1600)$ &1631\\
L=0&N=1 & 70  &$^28$   &                    &1530& $^210$ & $\Delta_{1/2^+}(1750)$ &1631\\
L=0&N=2 & 56  & $^28$  & N$_{1/2^+}(1710)$   &1779& $^410$ & $\Delta_{3/2^+}(1920)$ &1950\\
L=0&N=3 & 56  & $^28$  & N$_{1/2^+}(2100)$     &2076& $^410$ && 2223 \\
\hline
L=0&N=0& 56  & $^28$  & $\Lambda_{1/2^+}(1115)$ &1143 &&& \\
L=0&N=1& 56  & $^28$  & $\Lambda_{1/2^+}(1600)$ &1565 &&& \\
L=0&N=2& 56  & $^28$  & $\Lambda_{1/2^+}(1810)$ &1895 &&& \\
\hline
L=0&N=0& 56  & $^28$  & $\Sigma_{1/2^+}(1193)$ & 1143 & $^410$ & $\Sigma_{3/2^+}(1385)$ & 1394 \\
L=0&N=1& 56  & $^28$  & $\Sigma_{1/2^+}(1560)$ & 1565 & $^410$ && 1757 \\
L=0&N=1& 70  & $^28$  & $\Sigma_{1/2^+}(1660)$ & 1664 & $^210$ & $\Sigma_{1/2^+}(1770)$ & 1757 \\
L=0&N=2& 56  & $^28$  & $\Sigma_{1/2^+}(1880)$ & 1895 & $^210$ && 2056     \\
\hline
L=0&N=0& 56  & $^28$  & $\Xi_{1/2^+}(1320)$    & 1317 & $^410$ & $\Xi_{3/2^+}(1530)$    & 1540 \\
L=0&N=1& 56  & $^28$  & $\Xi (1690)$    & 1696 & $^410$ && 1869 \\
\hline
L=0&N=0& 56  &        &                        &      & $^410$  & $\Omega_{3/2^+}(1672)$ &1672  \\
L=0&N=1& 56  &        &                        &      & $^410$  &  & 1984  \\
\hline
\end{tabular}
\ec
\renewcommand{\arraystretch}{1.0}
\end{table}
\subsection{Resonances in the first harmonic-oscillator band}
The lowest orbital-angular-momentum excitations have L=1; in SU(6) a
70-plet is expected which can be decomposed into SU(3) multiplets:
$$
70 = {^2}10\ \oplus\ {^4}8\ \oplus\ {^2}8\ \oplus\ {^2}1.
$$
\paragraph{The spin-3/2 resonances:}
Spin-3/2 plus L=1 forms a spin-triplet of
resonances, with quantum numbers
$1/2^-$, $3/2^-$, and $5/2^-$.  The three states 
N$_{1/2^-}(1650)$, N$_{3/2^-}(1700)$, N$_{5/2^-}(1675)$ obviously
match our expectation, as well as the 
$\Sigma_{1/2^-}(1750)$, $\Sigma_{3/2^-}(1670)$, $\Sigma_{5/2^-}(1775)$
resonances. (We remind the reader that the  $\Sigma_{3/2^-}(1670)$
could also be the partner of the  $\Sigma_{1/2^-}(1620)$, provided the
$\Sigma_{3/2^-}(1580)$ does not exist.) 
In the $\Lambda$ sector, the $\Lambda_{1/2^-}(1800)$
and $\Lambda_{5/2^-}(1830)$ can be assigned to the spin-3/2 octet
states, the  $\Lambda_{3/2^-}$ is missing.
\par
There is only one corresponding $\Xi$ resonance, the
$\Xi_{3/2^-}(1820)$. It is assigned to the 
octet but it could also belong to the decuplet where it should have a
mass of 1874 MeV. 
The masses of experimentally known
$\Omega$ resonances are incompatible with the calculated
masses of (L=1,N=0) or (L=0,N=1) resonances.
\par
We expect three spin-doublets: in SU(3)
singlet, octet, and decuplet.  
\paragraph{Singlet:} The two states $\Lambda_{1/2^-}(1405)$,
$\Lambda_{3/2^-}(1520)$ are very low in mass. From (\ref{mass})
we predict 1460 MeV for the singlet L=1 states, in reasonable 
agreement with the experimental values. The mass formula assumes that
instanton interactions lead to a mass shift in the singlet - with all
three quarks antisymmetric w.r.t. the exchange of two quarks - which is
three times larger than for the octet state where only one quark pair
is antisymmetric w.r.t. exchange of the two quarks.
\paragraph{Octet:} The states N$_{1/2^-}(1535)$, N$_{3/2^-}(1520)$ 
(expected mass 1530 MeV); $\Lambda_{1/2^-}(1670)$,
$\Lambda_{3/2^-}(1690)$ and $\Sigma_{1/2^-}(1620)$,
$\Sigma_{3/2^-}(1580)$ or  $\Sigma_{3/2^-}(1670)$ (expected mass 1664
MeV); and $\Xi_{3/2^-}(1820)$  (expected mass 1787 MeV) fill this slot; 
only the $\Xi_{1/2^-}$ is missing. 
\paragraph{Decuplet:} In case of $\Delta$ and $\Omega$ resonances, the
assignment is conceptually easy, even though there is no candidate for the two
$\Omega$ resonances. The $\Delta$ doublet is observed at 
$\Delta_{1/2^-}(1620)$, $\Delta_{3/2^-}(1700)$ and expected at 1631
MeV. 
\par
In case of the $\Sigma$ and $\Xi$, which contribute to the octet and
the decuplet, we expect two additional states. We have combined the two
states $\Sigma_{1/2^-}(1750)$, $\Sigma_{3/2^-}(1670)$ with the
$\Sigma_{5/2^-}(1775)$ to form a triplet of states, expected from the
$^48$ part of the 70-plet. Now, the two $^210$ states - expected
to have the same mass - are missing. 
Similarly, the $\Xi_{3/2^-}(1820)$ and the unobserved $\Xi_{1/2^-}$ 
could belong to the octet and to the decuplet. As octet states they
have spin 3/2, as decuplet states spin 1/2. In both cases, they do
not undergo instanton interactions and the predicted masses are the
same. These are {\em hidden} resonances in our nomenclature. 
Experimentally, there are indications for a doubling of states with
identical quantum numbers but different decay modes. See the
comment of the Particle Data Group in the full listing 
for $\Sigma (1670)$ and $\Sigma (1690)$ bumps.
\par
\subsection{Resonances in the second harmonic-oscillator band}
In the second band of the harmonic-oscillator the following multiplets
are expected:
\paragraph{$\bf (56,0^+_2)$ and $\bf (70,0^+_2)$:} The scalar excitations
were discussed in section \ref{scalar}. Both multiplets contain
entries. Possibly, there is a selection rule preventing 
$(70,0^+_2)$ states for three identical quark masses.
\paragraph{$\bf (20,1^+_2)$:} The multiplet has no component which is
antisymmetric in spin and flavor. The N$^*$ resonances are hidden behind 
the N$_{5/2^{+}}(2000)$ and N$_{3/2^{+}}(1900)$, one of
the $^41\ \Lambda$ states behind the $\Lambda_{5/2^{+}}(2020)$, the
other two  $\Lambda_{3/2^{+}}$ and $\Lambda_{1/2^{+}}$ are 
expected also at 2056 MeV. In these states, 
both harmonic oscillators are excited, each with one unit of angular 
momentum. (The two angular momenta couple to a total orbital angular 
momentum 1.) These states are difficult to excite but could be narrow. 
So far, we have no evidence that resonances are formed in $(20,1^+_2)$
and we do not discuss such states further down.
\paragraph{$\bf (56,2^+_2)$ and $\bf (70,2^+_2)$:} 
From the $(56,2^+_2)$
multiplet, we expect a spin-1/2 octet and a spin-3/2 decuplet. 
The octet states with L=2 and spin 1/2 
can be identified with the N$_{3/2^+}(1720)$ and N$_{5/2^+}(1680)$, 
the $\Lambda_{3/2^+}(1890)$ and $\Lambda_{5/2^+}(1820)$ 
and the $\Sigma_{3/2^+}(1840)$ and $\Sigma_{5/2^+}(1915)$. The two
states, $\Xi(1950)$ and $\Xi(2030)$, have the expected mass (2004 MeV)
but spin and parity are not known. 
Decuplet states with spin 3/2 are obviously the four
resonances $\Delta_{1/2^+}(1910)$, $\Delta_{3/2^+}(1920)$,
$\Delta_{5/2^+}(1905)$, and $\Delta_{7/2^+}(1950)$, all expected at
1950 MeV. The  $\Sigma_{7/2^+}$(2030) state 
which must have S=3/2; the $^2\Sigma_{3/2^+}(2080)$,
$^2\Sigma_{5/2^+}(2070)$ are natural partners in this super-multiplet.
The $\Xi (2120)$ and $\Xi (2250)$ have masses which are not
incompatible with the expected 2157 MeV. There is also the 
$\Omega (2470)$ resonance with a mass compatible with an L=2 excitation.
\par
The $\Sigma$ and $\Xi$ states assigned to the decuplet in 56-plet 
with spin 3/2 could
also belong to the octet with spin 3/2 in the 70-plet. As spin=3/2
states their masses are not affected by instanton-induced
interactions; these super-multiplets are predicted to coincide in
mass, the super-multiplets are {\em hidden}. The question if there
are states belonging to $^48$ multiplets can only be decided for
baryons which contribute only to the octet and not to the decuplet.
\par
There is one N$_{7/2^+}$ and one $\Lambda_{7/2^+}$ state at about 2
GeV; their masses are too low to assign an internal angular momentum
L=4 to these states. Then, they have to have spin S=3/2 and have to
belong to the 70-plet. These two states are very important; they
entail the existence of many further states. In the N sector, a
nearly complete quartet can be made up, the
N$_{3/2^+}(1900)$, N$_{5/2^+}(2000)$, N$_{7/2^+}(1990)$ 
(with the fourth member possibly seen by \cite{Plotzke:ua}). 
The $\Lambda_{5/2^+}(2110)$ is likely the companion of the
$\Lambda_{7/2^+}(2020)$. Hence there are L=2 states belonging to
the 70-plet. 
\par
In the 70-plet we expect not only the octet with spin 3/2
but also spin 1/2 multiplets, in singlet, octet and decuplet.  
As mentioned above, the decuplet-spin-1/2 states are 
mass-degenerate with the spin-3/2 states and are
hence difficult to establish. But the singlet and octet
states should be observable.
\par
We should expect a doublet of SU(3) singlet states with L=2
and S=1/2. These states undergo strong instanton interactions, with
a symmetry factor in eq. (\ref{mass}) of 1.5 instead of 1 (for
the octet $\Lambda$'s). The mass is calculated to 1809 MeV
while the octet states are expected at 1895 MeV. We note
that the $\Lambda_{5/2^+}(1820)$ fits excellently to the
mass prediction for the singlet state. However, to claim that the 
$\Lambda_{5/2^+}(1820)$ belongs to the singlet system - together 
with the $\Lambda_{1/2^-}(1405)$, $\Lambda_{3/2^-}(1520)$, and
$\Lambda_{7/2^-}(2100)$ - is certainly not justified without observing
a doublet structure. 
\par
The octet partners as nucleon excitation belonging to 
$^28$ in a 70-plet, with spin-parities N$_{3/2^+}$ and N$_{5/2^+}$,  
should have a mass of 1779 MeV but there are no known states. 
\begin{table}
\caption{Mass spectrum of observed and missing N resonances. 
Nominal masses are from the PDG listings (in MeV). The last column gives masses
from eq. (\protect\ref{mass}). 
}
\renewcommand{\arraystretch}{1.4}
\bc
\begin{tabular}{|lccc|c|c|} 
\hline
&&&&&M\\
\hline
L=2&N=0 & 56  & $^28$  & N$_{3/2^+}(1720)$, N$_{5/2^+}(1680)$ &1779\\             
L=2&N=0 & 56  & $^28$  & $\Lambda_{3/2^+}(1890)$, $\Lambda_{5/2^+}(1820)$ &1895\\
L=2&N=0 & 56  & $^28$  & $\Sigma_{3/2^+}(1840)$, $\Sigma_{5/2^+}(1915)$&1895\\
L=2&N=0 & 56  & $^28$  & $\Xi_{3/2^+}(xxx)$, $\Xi_{5/2^+}(1950)$ &2005\\
L=2&N=0 & 56  & $^410$ & $\Delta_{1/2^+}(1910)$, $^1\Delta_{3/2^+}(1920)$, $^1\Delta_{5/2^+}(1905)$,
$\Delta_{7/2^+}(1950)$ &1950\\
L=2&N=0 & 56  & $^410$ & $^2\Sigma_{1/2^+}(xxx)$,$^2\Sigma_{3/2^+}(2080)$, $^2\Sigma_{5/2^+}(2070)$,
$^2\Sigma_{7/2^+}(2030)$& 2056\\
L=2&N=0& 56  & $^410$  & $\Xi_{1/2^+}(xxx)$, $\Xi_{3/2^+}(xxx)$, $\Xi_{5/2^+}(xxx)$, $\Xi_{7/2^+}(2120)$ &2157\\
\hline
L=2&N=0 & 70  & $^28$  & N$_{3/2^+}(xxx)$, N$_{5/2^+}(xxx)$ &1866\\             
L=2&N=0 & 70  & $^48$  & N$_{1/2^+}(xxx)$, N$_{3/2^+}(1900)$, N$_{5/2^+}(2000)$, N$_{7/2^+}(1990)$ &1950\\             
L=2&N=0 & 70                & $^210$ & $^1\Delta_{3/2^+}(xxx)$,$^1\Delta_{5/2^+}(xxx)$ &1950\\       
L=2&N=0 & 70  & $^28$  & $\Lambda_{3/2^+}(xxx)$, $\Lambda_{5/2^+}(xxx)$ &1977\\             
L=2&N=0 & 70  & $^48$  & $\Lambda_{1/2^+}(xxx)$, $\Lambda (2000)$,
$\Lambda_{5/2^+}(2110)$, $\Lambda_{7/2^+}(2020)$ &2056\\
L=2&N=0 & 70  & $^28$ & $^2\Sigma_{3/2^+}(xxx)$, $^2\Sigma_{5/2^+}(xxx)$,& 1977\\
L=2&N=0 & 70  & $^48$ & $^2\Sigma_{1/2^+}(xxx)$,$^2\Sigma_{3/2^+}(xxx)$, $^2\Sigma_{5/2^+}(xxx)$,
$^2\Sigma_{7/2^+}(xxx)$& 2056\\
L=2&N=0 & 70  & $^210$ & $^2\Sigma_{3/2^+}(xxx)$, $^2\Sigma_{5/2^+}(xxx)$& 2056\\
\hline
L=2&N=0 & 70  &$^21$& $\Lambda_{3/2^+}(xxx)$, $\Lambda_{5/2^+}(xxx)$ &1809\\   
\hline
\end{tabular}
\ec
\renewcommand{\arraystretch}{1.0}
\end{table}

\subsection{Resonances in the third harmonic-oscillator band}
The third harmonic-oscillator band is in an intermediate range in which we
still observe some states which document the richness of the
three-particle dynamics, but the number of seen states is already
significantly reduced compared to our expectation. In particular
strange baryons are scarce and have mostly no spin-parity
determination. In the third band, we
expect the following multiplets:
\bc
$\rm (70,3^-)_{N=0}; (70,1^-)_{N=1}; (56,3^-)_{N=0}; (56,1^-)_{N=1}$. 
\ec
We do not necessarily expect two multiplets with the
same quantum numbers to lead to a duplication of states. 
In addition, there is a 70-plet with L=2 and even parity 
for which we have no
evidence. (They require both oscillators to be excited at the same
time.) Remember, the oscillator band is $N=3$ for L=1 and N=1.
\paragraph{$\bf (70,3^-)_{\bf\rm N=0}$:}
We find a doublet of nucleon resonances of negative parity, with J=5/2
and 7/2, at 2190 and 2200 MeV, respectively, which match our
expectation for the  $\rm ^28(70,3^-)_{N=0}$ multiplet. The
N$_{9/2^-}(2250)$ has intrinsic spin S=3/2 and is assigned to the
$\rm ^48(70,3^-)_{N=0}$ multiplet. In the $\Lambda$ sector
we should expect two $\Lambda_{7/2^-}$ and two
$\Lambda_{5/2^-}$, one doublet at 2100 MeV due to the singlet system,
one doublet at 2248 MeV belonging to the octet. Only one state is
known, the $\Lambda_{7/2^-}(2100)$ which we interpret as singlet
state. The  $\Sigma_{7/2^-}(2100)$ with its rather low
mass weakens somehow the evidence for the assignment
of the $\Lambda_{7/2^-}(2100)$ as SU(3) singlet state. The 
$\Sigma_{7/2^-}(2100)$ is however a 1* resonance only; 
the  $\Lambda_{7/2^-}(2100)$ with its 4*
is certainly much more reliable. The $\Sigma (2245)$ - with
unknown spin-parity - fits excellently to L=3 spin-1/2 octet 
hypothesis. One $\Xi$ state, the $\Xi(2370)$, may have L=3 when
the mass formula is used as a guide.
The $\Delta_{7/2^-}$(2200) is a natural case to be assigned to 
$\rm ^210(70,3^-)_{N=0}$. 
Hence we have evidence for all parts
of the SU(3) decomposition of the  $\rm (70,3^-)_{N=0}$
super-multiplet, of the $^210, ^28, ^48$ and $^21$ multiplets.  
\paragraph{$\bf (56,3^-)_{\bf\rm N=0}$:}
The $(56,3^-)_{N=0}$ super-multiplet requires the existence
of a $\Delta_{9/2^-}$ resonance at 2223 MeV. There is no evidence for
such a state. At L=1, S=3/2 decuplet states are forbidden; we assume
that this selection also holds for higher odd-orbital-angular-momentum
resonances. Nucleon resonances with spin 1/2 do exist; they can
however also be assigned to the $\rm (70,3^-)_{N=0}$ as discussed
above. Hence we do not find evidence that the
$\rm (56,3^-)_{N=0}$ super-multiplet is needed dynamically.
\paragraph{$\bf (56,1^-)_{\bf\rm N=1}$:}
Resonances with $\rm (56,1^-)_{N=1}$ exist; the three 
states $\Delta_{5/2^-}(1930)$, $\Delta_{3/2^-}(1940)$, and 
$\Delta_{1/2^-}(1900)$ are likely members of this super-multiplet. 
\paragraph{$\bf (70,1^-)_{\bf\rm N=1}$:}
We now turn to the $(70,1^-)_{\rm N=1}$ resonances. 
N$^*$'s belonging to the $(70,1^-)_{\rm N=1}$ super-multiplet are
missing in the Tables; they should have a mass of 1866 MeV. 
The SAPHIR Collaboration has suggested two states which would fit
to this prediction, a N$_{1/2^-}(1897)$ \cite{Plotzke:ua} observed
in the  N$\eta^{\prime}$ decay mode and a N$_{3/2^-}$(1895) 
\cite{Bennhold:1997mg} decaying into K$^+\Lambda$. 
$\Delta^*$'s in the $(70,1^-)_{\rm N=1}$ super-multiplet 
should have a mass of 1950 MeV. The 
three states $\Delta_{5/2^-}(1930)$, $\Delta_{3/2^-}(1940)$, and
$\Delta_{1/2^-}(1900)$ are naturally assigned to the $\rm
(56,1^-)_{N=1}$ super-multiplet; the two states $\Delta_{3/2^-}(1940)$ and
$\Delta_{1/2^-}(1900)$ could also come from the
$(70,1^-)_{\rm N=1}$ super-multiplet; according to the mass formula
there is double-occupation, two states are hidden.
\par
No $\Lambda$ resonances are known which could be ascribed to 
the $(70,1^-)_{\rm N=1}$ super-multiplet, but there are
the two states $\Sigma_{1/2^-}(2000)$ and $\Sigma_{3/2^-}(1940)$ 
which are predicted to have a mass of 1977 MeV.  
\par
In summary, in the third harmonic oscillator band we find evidence for
the three super-multiplets $\rm (70,3^-)_{N=0}$, $\rm (56,1^-)_{N=1}$
and  $\rm (70,3^-)_{N=1}$, while the $\rm (56,1^-)_{N=0}$ and the
20-plets are missing.

\subsection{High-mass resonances}
At high masses, essentially only N$^*$'s and $\Delta^*$'s with high
total angular momenta are known. For even parity, nucleons 
are in a 56-plet: there are the N$_{9/2^+}(2220)$ and
the N$_{13/2^+}(2700)$ but there are no known N$_{11/2^+}$ or N$_{15/2^+}$  
with S=3/2 and L=4 or 6, respectively. 
There is only one negative-parity nucleon with orbital angular
momentum L=5, the N$_{11/2^{-}}(2600)$ which is likely a 
$^28\rm (70,L^-)_{N=0}$ resonance. We mention that 
the two states N$_{1/2^-}(2090)$ and N$_{3/2^-}(2080)$ can be
interpreted as second radial excitations.
\par
The $\Delta$
resonances with even parity are naturally given S=3/2 - as expected
for decuplet states in the 56-plet: the $\Delta_{11/2^+}(2420)$ 
and $\Delta_{15/2^+}(2950)$ evidence that the intrinsic spin is
S=3/2. The $\Delta_{11/2^+}(2420)$ is accompanied by the
$\Delta_{7/2^+}(2390)$ and $\Delta_{9/2^+}(2300)$. We assume they 
have orbital angular momentum L=4, coupling to J=5/2 (missing),
J=7/2, 9/2 and 11/2. Negative-parity $\Delta$ resonances also 
like to have S=3/2: there are the $\Delta_{9/2^-}(2400)$ and the
$\Delta_{13/2^-}(2750)$ which should - based on the Regge trajectories -
have intrinsic orbital angular momenta 3 and 5, respectively. In the
first and third harmonic oscillator levels, $\Delta$'s without
additional radial excitation (N=0) should be in a 70-plet and have
spin 1/2. Therefore we assume that the latter two states also have 
one quantum of radial excitation energy (N=1). The calculated masses
support this conjecture. L=3 and S=3/2 couple to 3/2, 5/2, 7/2 and
9/2. The  $\Delta_{5/2^-}(2350)$ has the right mass and quantum
numbers to fall into this super-multiplet, together with the
$\Delta_{9/2^-}(2400)$. The $\Delta_{1/2^-}(2150)$
could be a second radial excitation, like the  N$_{1/2^-}(2090)$ and
N$_{3/2^-}(2080)$.  
\par 
In summary, we observe N$^*$ and $\Delta^*$ resonances with
even parity as members of a 56-plet. Odd-parity nucleon
resonances are observed in $^28\rm (70,L^-)_{N=0}$ multiplets, 
$\Delta$'s in the  $^410\rm (56,L^-)_{N=1}$ multiplets. 
The reason for these preferences are unknown; we remind the reader
that not all solutions of the Hamiltonian need to be realized
in nature as stable rotations. 
The assignments of high-mass $\Sigma$, $\Xi$ and
$\Omega$ states given in the Tables (for which spin-parities are not
measured) are based on these selection rules.
\subsection{Suggestions for further experiments}
A first point of experimental interest would be to confirm the
reliability of the mass formula. Both, the quark model using
one-gluon-exchange \cite{Capstick:bm} and the quark model with
instanton-induced interactions \cite{Metsch} predict the three states
$\Delta_{5/2^-}(1930)$, $\Delta_{3/2^-}(1940)$, and 
$\Delta_{1/2^-}(1900)$ at masses in the 2200 MeV region. It has been
proposed \cite{cbelsa}
to search for the $\Delta_{3/2^-}(1940)$ in its hypothetical 
$\Delta_{3/2^-}(1940)\to\Delta_{3/2^+}(1232)\eta$ decay. A
confirmation would validate the model proposed here and would be
difficult to incorporate in those two quark models. 
\par
The  $\rm (70,1^-)_{N=1}$ multiplet predicts nucleon resonances with
spin 1/2 at 1866 MeV. The SAPHIR collaboration suggested a N$_{1/2^-}$
resonance at 1897 MeV \cite{Plotzke:ua}. The resonance was observed below
the N$\eta^{\prime}$ threshold in the  N$\eta^{\prime}$ decay
mode. A further resonance is claimed, a  N$_{3/2^-}$, at 1895 MeV in
its decay into K$^+\Sigma^0$ \cite{Bennhold:1997mg}. The two resonances are
perfectly compatible with forming  part of the  $\rm (70,1^-)_{N=1}$
multiplet. A high-statistics study including polarization observables
should decide if these claims are
justified. 
\par
The model predicts a further positive-parity 
N$^*$ doublet with J=5/2 and 3/2 between the low mass states at 1700
MeV and the states at 1950 MeV. These states should be observable
at Jlab and at ELSA in two-pion photo- or electroproduction
experiments \cite{Napolitano:1993kf,Thoma-prop}. The 1900 to 2000 MeV
region hosts a large number of N$^*$ and $\Delta^*$ states and further
resonances are expected to hide in this mass range. A very detailed
study of several final states is needed to test this prediction.
\par
It would be highly interesting to explore the high-mass range to see
if the selection rules are artifacts of the analyses which identify
most easily the highest partial wave, or if these selection rules
really show how a high-spin baryon adjusts its shape and internal
degrees of freedom to the large centrifugal forces. 
\par
\section{Interpretation}
\subsection{A new interpretation of strong QCD}
Given the simplicity of the mass formula, the agreement between  
experimental values and the prediction is remarkable. Hence we have to
discuss what the reason for this good agreement might be. 
At the first glance there is a clear contradiction. All baryon
resonances are rather well reproduced with a model which 
takes the string constant and the radial excitation energy from
mesons, from a quark-antiquark system. Obviously baryon 
resonances behave like quark-diquark excitations. 
\par
Diquark models of baryons have been suggested frequently but they 
have an intrinsic weakness. Let us consider the lowest orbital angular
momentum excitation, the N$_{3/2^-}(1520)$D$_{13}$ resonance. 
The intrinsic orbital angular momentum is 1, it combines with the
intrinsic spin 1/2 to a total angular momentum J=3/2. Now, not only
one quark is excited to an orbital angular momentum 1, both harmonic 
oscillators are coherently excited. This splitting of the oscillatory
motion is required by Fermi-Dirac statistics, by the requirement that
the wave function should be antisymmetric with respect to the exchange
of two quarks. Diquark models usually circumvent this demand by
assuming that diquarks are tightly bound so that anti-symmetrization of
quark pairs, one outside and one inside of the diquark, is not
required. Experimentally there is no support for a tightly bound diquark;  a
diquark model is also not consistent with the forces expected from
QCD. 

There is one way out: We may assume that baryon resonances are
quark-diquark excitations in color space. So we assume that the
dynamics is given by color. Antisymmetrization can be arranged by
flavor exchange. So a red up-quark may turn into a red down-quark. It
maintains color and exchanges its flavor. Now we have to explain why
color exchange is a slow process.

We propose a somehow unusual concept for constituent
quarks. We assume that the strong color-forces polarize the quark
and gluon condensates of the QCD vacuum. The current quark
plus its polarization cloud forms what we call now a constituent quark
of defined color. Color exchange is screened by the polarization
cloud. When a gluon is emitted it is re-absorbed in the polarization
cloud. Color propagates only stochastically from one color center to
another center within a polarization cluster, globally the
constituent quark keeps its color for a finite time, which is longer
than the life time for flavor exchange and comparable to the life
time of the baryon resonance. Flavor is not a property of a
constituent quark. The matrix element governing color exchange is not
known; we estimate it to be in the order of $\Lambda_{\rm QCD}$. 
\par 
In contrast to color exchange there is a fast flavor
exchange. Flavor exchange is not shielded by the polarised
condensates; flavor propagates freely in the QCD vacuum. 
Flavor exchange is possible via long-range meson-exchange or by 
instanton interactions at the surface of two neighboring colored
constituent quarks. Flavor exchange acts at a time 
scale given by the constant responsible for chiral symmetry breaking, 
$\Lambda_{\chi}$. In this picture confinement originates from
Pomeron-exchange-like forces transmitted by the polarization of the
vacuum condensates. Different SU(6) multiplets with the same
quantum numbers differ in their internal spin-flavor structure
which has no significance for the color wave function, and thus
no importance for the mass. 
\par
Color is usually not considered to be an observable quantity. For
the baryon ground states, this assumption is true. All three quarks
are in the same phase-space cell, the quarks cannot be localized and
their individual properties are not observables. This argument is not
applicable to excited baryons. If a baryon has an intrinsic orbital
angular momentum of, let us say, L=4, one quark is well separated in
phase space from the two other quarks. Its color can be defined and
can be shielded by its self-generated polarization cloud.
\par
This picture suggests that the largest contribution to the mass of a
hadron comes from the mass density of the polarization cloud and the
hadronic volume. This idea can be tested in a string model of
quark-diquark interactions. We assume that the polarization cloud 
between quark and diquark is concentrated in a rotating flux tube
or a rotating string with a homogeneous mass density. The length of
the flux tube is $\rm 2r_0$, its transverse radius R. The velocity at 
the ends may be the velocity of light. Then the total mass of the
string is given by \cite{Nambu:1978bd}
\begin{equation}
\rm Mc^2 = 2\int_{0}^{r_0}\frac{k dr}{\sqrt{1-v^2/c^2}} = kr_0\pi
\end{equation}
and the angular momentum by
\begin{equation}
\rm L = \frac{2}{\hbar c^2}\int_{0}^{r_0}\frac{k r v dr}{\sqrt{1-v^2/c^2}} = 
\frac{k r_{0}^{2} \pi}{2\hbar c}
\end{equation}
The orbital angular momentum is proportional to 
\begin{equation}
\rm L = \frac{1}{2\pi k\hbar c} M^2
\end{equation}
From the slope in Fig.~\ref{N-Delta} we find $k = 0.2$\,GeV$^2$
and a radius of 
\begin{equation}
\rm  r_0\left(\Delta_{15/2^+}(2950)\right) = 4 {\rm fm} 
\end{equation}
According to the Nambu model, excited quark and the diquark 
in the $\Delta_{15/2^+}(2950)$ are separated by 8\,fm\,!  
\par
The volume of the flux tube is 2$\rm\pi R^2r_0$, the mass density
\begin{equation}
\rho = \frac{k}{2R^2c^2}.
\end{equation}
We now assume that the mass density in the $\Delta (1232)$ is the same
as the one in the flux tube. We thus relate
\begin{equation}
\rm \frac{4}{3}\pi R^3\cdot\rho  = M_{\Delta (1232)}
\end{equation} 
which gives a radius of the polarization cloud of the $\Delta (1232)$
of 0.6\,fm (and 0.37fm for the $\rho$). This is not unreasonable, even
though smaller than the RMS charge radius of the proton. However,
an additional pion cloud may increase the charge radius. 
\subsection{Consequences of the colored-constituent-quark concept}
The assumption that constituent quarks have a defined color, and that
color exchange is shielded by the polarization cloud offers 
a new interpretation for a large number of
phenomena which are not yet understood.
\paragraph{Confinement:}
When two quarks are separated, the volume in which the QCD vacuum is
polarized increases 
with the quark-quark separation. The net color charge does not
change, hence the energy stored in the polarized condensates 
increases linearly: the confinement potential is a linear function of
the quark separation.
This interpretation of the confinement potential 
follows immediately from the assumption that color exchange between
two quarks is a slow process and this in turn is the consequence of
the similarity of the meson and of the baryon string tension. 
\paragraph{Structure functions:}
The polarization clouds surrounding the current quarks are of course
seen in deep inelastic scattering, the quarks directly, the gluons
through their contribution to the total momentum.   
\paragraph{The spin crisis:}
It was a surprise when it was discovered that the
proton spin is not carried by quarks. The success of the naive quark
model in the prediction of the ratios of magnetic moments of octet
baryons seemed to be a solid basis for the assumption that the spin of
the proton should be carried by its 3 valence quarks. But this
naive expectation fails, 
the contribution of all quark- and antiquark-spins to the proton spin
is rather small \cite{Filippone:2001ux}. A large fraction from the proton spin
must be carried by the intrinsic orbital angular momenta of quarks or
by orbital or spin contributions of gluons. We assume that the
magnetic moment of the spin induces polarization into the condensates. The
polarized gluon condensates provide a gluonic contribution to the
proton spin, the quark condensate a spin and orbital contribution to
it. Orbital angular momenta of quarks enter because the quarks in the
condensate are pairwise in the $^3{\rm P}_0$ state. The orientation
defined by direction of the current-quark spin may induce
internal currents which contribute to the magnetic moment. 

An analogy can be found in superconductivity. If a magnet moment is
implanted into a superconducting material, the superconductivity may
be destroyed. If it is maintained the Cooper pairs will be polarized
and the currents adjust to take over part of the magnetic moment of
the alien element. 

\paragraph{The $^3{\bf\rm P}_0$ model:} 
A further example for the usefulness of the concept proposed here
is the $^3{\rm P}_0$ model for meson and baryon decays. According to
this model the quantum numbers of a $\rm q\bar q$ pair, created in 
a decay process, have the quantum numbers of the vacuum. 
These quantum numbers are preserved, when a $\rm q\bar q$
pair from the condensate is shifted to the mass shell. 
\paragraph{Hybrids and glueballs:}
Finally, the picture also provides a new interpretation of glueballs 
and hybrids. In the flux tube model hybrids are thought of as
quark-antiquark pairs connected via a gluonic string forming a flux tube. 
This flux tube can be exited and such excitations are called hybrids
if they are resonant. Here the question arises if the polarization
status of the gluon and quark condensates can undergo oscillatory
motions.  This is a dynamical question and
not only a question of mean forces and of an energy minimum. There is,
e.g., the possibility that the polarization cloud supports 
longitudinal oscillations, sound-like waves. These can be identified
with radial excitations. Their existence is well established; this
does not imply that the polarization cloud also supports transverse
oscillations. We emphasize that in the baryon spectrum, there is no
evidence for additional states, neither hybrid states (with the flux
tube excited) nor for penta-quarks.
\par
Glueballs can be seen as polarization clouds without a color charge
driving the polarization. Here, the question is if the gluon and quark
condensates support soliton-like solutions which propagate in free
space and which manifest themselves as energy bumps. A recent critical
review on the status of hybrids and of the scalar glueball can be
found in \cite{Klempt:2000ud}.
\paragraph{The Color-Constituent-Quark model and lattice QCD}
The problems which are discussed here seem unsolvable
within perturbative QCD or in quenched lattice calculations. 
The effect a color charge induces in the quark   
and gluon condensates are not within the range of present-day lattice QCD. 
Even the inclusion of a small number of virtual $\rm q\bar q$ pairs is
unlikely to represent the full complexity 
of the QCD vacuum and its response to a color source.  
\section{Summary}
We have proposed a new baryon mass formula which reproduces with good
precision most of the available baryon masses. The formula is based on
the observation that the slope of mesonic and baryonic Regge
trajectories are the same, that radial excitations of mesons and
baryons have similar level spacings and 
 that mass splittings ascribed to one-gluon exchange
like color-magnetic spin-spin and spin-orbit interactions do not play  
an important role for baryon masses. The
octet-decuplet mass splitting is shown to depend on the symmetry of
the wave function; the splitting is not induced by spin-spin color-magnetic
interactions but by instantons. The richness of the baryon spectrum
prevents an interpretation of baryon resonances as quark-diquark
excitations. We propose therefore that the masses are given by
the dynamics of colored quark clusters, constituent quarks of defined
color. Fast flavor exchange fulfills the requirement of Fermi-Dirac
symmetry. 

In our view, 
the three quarks of a baryon polarize the vacuum condensates which
then shield the color charge preventing a fast color exchange. As
color exchange is slow, there is no color-magnetic interaction. 
Baryon masses depend on the dynamics of colored constituent quarks or
colored quark clusters. Flavor is not a property of constituent
quarks. 

This view of constituent quarks as colored quark clusters has
wide-reaching consequences which are briefly outlined. The view offers
a language in which confinement can be described and the proton spin
crisis be qualitatively understood. It offers new interpretations for
so-called gluonic excitation modes in hadron spectroscopy, of hybrids
and glueballs. We are convinced that baryons provide key issues for an
improved understanding of QCD in the confinement region.

\subsection*{Acknowledgments}
I would like to thank D. Diakonov, K. Goeke, U. L\"oring, D. Merten
B. Metsch, B. Schoch, Chr. Weinheimer for discussions on various
parts of this paper, and J. Lutz for help in the preparation of
the Figures.

\end{document}